\documentstyle[epsfig,fleqn,leqno]{article}
\oddsidemargin---2mm
\newcommand{\HU}{\hat{U}}
\newcommand{\HG}{\hat{G}_\lambda}
\newcommand{\HH}{\hat{H}}
\newcommand{\IR}{\hbox{{\rm I}\kern-.2em\hbox{\rm R}}}
\newcommand{\balpha}{ \mbox{\boldmath$\alpha$}}
\newcommand{\bgamma}{ \mbox{\boldmath$\gamma$}}
\newcommand{\bphi}{ \mbox{\boldmath$\phi$}}
\newcommand{\btheta}{ \mbox{\boldmath$\theta$}}
\newcommand{\bvarphi}{ \mbox{\boldmath$\varphi$}}
\newcommand{\bzeta}{ \mbox{\boldmath$\zeta$}}
\newcommand{\brho}{ \mbox{\boldmath$\rho$}}
\newcommand{\bw}{ \mbox{\boldmath$w$}}
\newcommand{\bp}{ \mbox{\boldmath$p$}}
\newcommand{\bq}{ \mbox{\boldmath$q$}}
\newcommand{\bk}{ \mbox{\boldmath$k$}}
\newcommand{\bl}{ \mbox{\boldmath$l$}}
\newcommand{\bm}{ \mbox{\boldmath$m$}}
\newcommand{\bn}{ \mbox{\boldmath$n$}}
\newcommand{\be}{\begin{equation}}
\newcommand{\ee}{\end{equation}}

\begin{document}%

%%%%%%%%%%%%%%%Title page%%%%%%%%%%%%%%%%
\title{\Large \bf
Exact evolution operator on non-compact group manifolds}
\author{
{\normalsize\sc Nurit Krausz and M. S. Marinov
%\thanks{\quad Supported by G. I. F.}
}\\
{\normalsize\sl
Department of Physics, Technion-Israel Institute of Technology}\\
{\normalsize\it Haifa 32000, Israel}
%{\normalsize\it
%Max-Planck-Institut f\"{u}r Physik and Astrophysik}\\
%{\normalsize\it Werner-Heisenberg-Institut, M\"{u}nchen, Germany }\\
}

\date{}  

\maketitle
%%%%%%%%%%%%%%%%%%%%%%%%%%%%%%%%%%%%%%%%%%%%%%
{\begin{abstract}
Free quantal motion on group manifolds is considered. The Hamiltonian
is given by the Laplace -- Beltrami operator on the group
manifold, and the purpose is to get the (Feynman's) evolution kernel $K_t$.
The spectral expansion, which produced a series of the representation
characters for  $K_t$ in the compact case, does not
exist for non-compact group, where the spectrum is not bounded.
In this work real analytical groups are investigated, some of which
are of interest for physics. 
An integral representation for $K_t$ is obtained in terms of the Green
function, i.e. the solution to the Helmholz equation on the group
manifold. The alternative series expressions for the evolution operator
are reconstructed from the same integral representation, 
the spectral expansion (when exists) and the sum over classical paths.
For non-compact groups, the latter can be interpreted as the ({\em exact})
semi-classical approximation, like in the compact case.
The explicit form of $K_t$ is obtained for a number of non-compact groups.

 \end{abstract}

\section{Introduction}
Normal physical systems have energy spectra bounded from below, so a
stable ground state exists and may be considered as a zero-temperature
limit of the Gibbs thermal state. If the Hamiltonian operator $\HH$ is
positive semi-definite (we shall use the notation $\HH\ge 0$), the Gibbs
density operator $\exp(-\beta\HH)$ exists and its kernel is the
fundamental solution of the Bloch equation
$\partial\Psi/\partial\beta=-\HH\Psi$, for $\beta>0$. This equation,
however, has no stable solution if the spectrum of $\HH$ is
extended to $-\infty$.

On the other hand, the Schr\"{o}dinger equation and the corresponding
evolution operator $\HU_t\equiv\exp(-it\HH)$,
\begin{equation}%(1)
i\partial\Psi/\partial t=\HH\Psi,\;\;\;\;
\Psi_t=\HU_t\Psi_0,
\end{equation}
may be meaningful for a regular self-adjoint operator $\HH$
even if its spectrum is not bounded at all. The evolution operator
may be defined for any real $t$ (in a properly define Hilbert space of the
wave functions $\Psi$), even if it cannot be continued analytically to the
complex (lower-half) $t$-plane, as in the usual case, where points on the
negative imaginary axis correspond to inverse temperatures. Note that
(1)  is a wave equation for real $t$ (even being first-order in $t$),
and not a parabolic heat-transport (or Bloch) equation, where the positive
definiteness of $\HH$ is essential. Actually, it may be extended to the
real form $\partial^2\Psi/\partial t^2+\HH^2\Psi=0$, which is of the
hyperbolic type, since $\HH^2\ge 0$ (like $-\Delta$ in the standard wave
equation), so the Cauchy (initial-value) problem has a proper solution
\cite{hadamard}.

It is remarkable that equations of the type (1) with non-definite
$\HH$ were considered extensively for functions on the pseudo - Euclidean
(Minkowski) space. In that case, $\HH=\Box$, i.e. the d'Alembertian (or a
more complicated operator in presence of an external field), and $t$
played the role of the {\em proper time} (the classical references are
\cite{fock,nambu,schwinger,stuckelberg}). 

Analysis of the Schr\"{o}dinger-type equation (1) for a (non-definite)
operator $\HH$ enables one to get an insight into the properties of its
spectrum and the eigen-functions, as was the case in the proper-time
formalism. We shall consider the free quantal motion on group manifolds,
which is described by Eq. (1) where $\HH=-\Delta$ is the second-order
Laplace -- Beltrami operator. For compact groups, $-\Delta\ge 0$, and
the complete solution is known for decades (see Section 2 for references).
As to real non-compact groups, $\Delta$ is well defined but indefinite,
like $\Box$ in the Minkowski space. It is an interesting class of
problems which can be also solved completely, as shown in this work.

If $\HH\ge 0$, the evolution operator can be represented by its spectral
expansion, which is a convergent series for $t > 0$ (and $\HU_0=\hat{I}$
-- the unit operator),
\begin{equation}%(2)
\HU_t=\sum_{n\ge 0}e^{-i\varepsilon_nt}\psi_n\otimes\psi_n^*,\;\;\;
\HH\psi_n=\varepsilon_n\psi_n.
\end{equation}
In the situation concerned here, the series would not be convergent, 
yet one can use the Laplace representation of $\HU_t$ in terms of the
resolvent $\HG$,
\begin{equation}%(3)
\HU_t=\frac{1}{2\pi i}\int_{\sf C}\HG e^{-i\lambda t}d\lambda,
\;\;\;\HG \equiv(\HH-\lambda)^{-1}.
\end{equation}
The contour $\sf C$ in the complex $\lambda$-plane should be defined
properly, with account of singularities of $\HG$, which take place
on the real axis, since $\HH=\HH^\dagger$. It is assumed that the contour
is in the upper half-plane, according to the principle of causality, so
that $\HU_t=0$ for $t<0$.

Now the resolvent generates the solution to the inhomogeneous Helmholz -
type equation \cite{morse}, within the properly defined class of
functions,
\begin{equation}%(4)
(\HH-\lambda)\psi=f,\;\;\;\psi_\lambda=\HG f.
\end{equation}
(Note that for $\HH=\Box$ and $\lambda=-(m^2-i\epsilon)$, $\epsilon>0$,
the integral kernel of $\HG$ is just the standard causal propagator of a
massive scalar particle.) For $\HH\ge 0$, the singularities of $\HG$ are
all on the half-axis Re$\lambda\ge 0$. In that case, for $t>0$, the
contour $\sf C$ may be deformed to the lower half-plane and closed there 
at $\infty$ (fig. 1), so the residues of the poles at $\lambda=\varepsilon_n$ 
(and the imaginary part of Green's function for the continuous part
of the spectrum, respectively) would reproduce the spectral expansion (2).
In the case concerned here, however, singularities appear on the whole
real axis, and Eq. (3) provides with a more general representation which 
cannot be reduced to the series in (2). Now equation (4) is of the
hyperbolic type, like the Klein -- Gordon equation, yet it should be
solved for all (complex) values of $\lambda$ which appear in the Laplace
integral (3).

The problem is solved in the following way.
The Helmholz equation on the group manifold is considered, and the
integral kernel of the resolvent operator is constructed explicitly. 
The result is the integral representation (3) for the evolution kernel,
to be subject to a further analysis.
Remarkably, the integral for the evolution kernel can be represented as a
series of terms which correspond to contributions from ``classical paths"
(geodetics) connecting the points of the group manifold to its origin (the
group unity). For non-compact groups, like in the compact case, the
semi-classical approach (including the pre-exponential factor) leads to
the {\em exact} result, provided that {\em all} classical paths are taken
into account. For compact groups the series are infinite and $r$-fold 
($r$ is the group rank). In contrast, each non-compact group is splitted
in a number of classes (like the Minkowski space, containing ``space-like"
and ``time-like" vectors). Each class has its specific set of paths and
the corresponding series for the evolution kernel.

After a description of the approach in general in Sections 2 and 3, 
the known results are reproduced for compact groups in section 4.
In section 5 the evolution operator is built for non-compact groups, but
the expression that is obtained depends on the maximal torus topology,
which is different in different domains on the manifold, and corresponds to
the different `classes' of the classical paths.  
The mathematical tools that are needed in order to analyze non-compact groups, 
and in particular the different domains and the corresponding maximal tori are given in 
section 6.
Some particular examples of real non-compact groups are presented
in Sections 7 and 8. Necessary mathematical notations and results are
given in Appendix.

 \section{Free motion on a group manifold} 
The Lie groups have the natural Riemannian structure given by the Cartan
-- Killing metrics (see Appendix A). For semi-simple Lie groups, 
the Riemannian metric is non-degenerate, and the invariant second-order 
 Laplace -- Beltrami operator (called here Laplacean $\Delta$) is defined,
as usual. (A general theory of invariant differential operators on the
group manifolds was given by Berezin\cite{ber56,ber57}.) 
Free motion on the group manifold is introduced naturally by means of  
the Schr\"{o}dinger-type wave equation,
 %\begin{eqnarray}
    \begin{equation}
\label{ka}
i\partial\Psi(g)/\partial t=-\Delta\Psi(g),\;\;\;
\int_{\sf G}|\Psi(g)|^2d\mu(g)=1,
    \end{equation}
%\end{eqnarray}
where $\Psi(g)$ is a (square-integrable) function on the group, $g\in{\sf
G}$, and $d\mu(g)$ is the invariant measure on ${\sf G}$, given by the
Riemannian structure.
The solution of the wave equation (\ref{ka}), for any initial condition 
$\Psi_0(g)$, is given by the evolution kernel ${\cal K}_t$,
       \begin{equation}
\label{kb}
\Psi_t(g_t)=\int_{\sf G}{\cal K}_t(g_t,g_0)\Psi_0(g_0)d\mu(g_0),\;\;\;
{\cal K}_t(g_t,g_0)\equiv \langle g_t|\HU_t|g_0\rangle.
    \end{equation}
Because of the invariance of Eq. (\ref{ka}) under the shifts on the group,
$g\rightarrow g_1gg_2$ for all $g_1,g_2\in{\sf G}$, the evolution kernel 
is reduced to an invariant function on the group manifold,
%   \begin{eqnarray}
     \begin{equation}
\label{kc}
{\cal K}_t(g_t,g_0)\equiv K(g_tg_0^{-1});\;\;\;
K_t(g)=K_t(g_1gg_1^{-1}),\;\;\;\forall g_1\in{\sf G}. 
   \end{equation}
%   \end{eqnarray}
Moreover, because of the latter property, the evolution kernel $K_t(g)$ 
is a {\em central function}. Namely, it
depends in fact only on the element of the Cartan subgroup, i.e. the 
maximal torus $\sf T$, $h\in{\sf T\subset G}$, where $g=vhv^{-1}$. The 
element $v$ is a representative of the coset space ${\sf V=G/T}$,
and $K_t(g)\equiv K_t(h)$ is independent of $v$.

The wave equation on the group $SU(2)$ was considered by Bopp 
and Haag\cite{bopp} and Schulman\cite{schulman}. 
Schulman presented the explicit solution for $SU(2)$
as well as for $SO(3)=SU(2)/Z_2$ and showed that the semi-classical
approximation is exact in that case. The heat transport equation,
which may be considered as the analytical 
continuation of Eq. (\ref{ka}) to negative imaginary values of $t$,
was solved by Eskin\cite{eskin} for all compact Lie groups.
The solutions of the wave equation for {\em compact} groups were
considered in a number of works\cite{dowker70,dowker71,marter78,mlong};
see also Ref.\cite{camporesi} for a review.

For compact Lie groups, the evolution kernel has the {\em spectral
expansion}, which is the sum over all unitary irreducible representations,
   \begin{equation}
\label{spex}
K_t(h)=\frac{1}{V_{{\sf G}}}\sum_{{\bl}\in\Lambda}d_{\bl}\chi_{\bl}(h)
\exp(-i\lambda_{\bl} t).
    \end{equation}
Here ${\bl}$ are the representation dominant weights, $\Lambda$ is the 
weight lattice in the $r$-dimensional root space ($r=$ rank$({\sf G})$),
$d_{\bl}$ is the representation dimensionality, $\chi_{\bl}(h)$ is the
representation character, $\lambda_{\bl}$ is the eigen-value of the
second-order Casimir operator, corresponding to the Laplacean $-\Delta$,
and 
   \begin{equation}   %(9)
\label{volume}
 V_{{\sf G}}\equiv\int_{\sf G}d\mu(g)
    \end{equation}
is the invariant volume of the group manifold. 

On the other hand, employing the Poisson transformation for $\theta$
-functions\cite{bellman}, the evolution kernel
can be represented as a sum over all the {\em classical trajectories} 
(geodetics) on the group manifold, connecting $g_0$ and $g_t$. 
The geodetics are described by means of a $r$-dimensional vector 
$\bvarphi$ in the Euclidean space tangent to ${\sf T}$. The result is
       \begin{equation}
\label{scap}
K_t(g)=\frac{1}{(4\pi it)^{n/2}}
\sum_{\bm\in\Gamma}F_{\bm}(\bvarphi)
\exp\left[i\frac{S_{\bm}^2(\bvarphi)}{4t}+it\frac{n}{24}\right].
          \end{equation}
Here $n$ is the group dimensionality, $\bm$ is the winding-number 
vector on the lattice $\Gamma$, dual to $\Lambda$, $S_{\bm}(\bvarphi)$ is
the distance from the origin to the point $h\in{\sf T}$, as measured along a 
line winded a number of times around the torus, and $F_{\bm}(\bvarphi)$ is
a known function (see in Section 4), which appears as the van Vleck
determinant of the semi-classical approximation and may be considered as
the first quantum correction. The first term in the exponent is the
classical action, and the second term is the constant (and the last)
quantum correction. It was shown that for any compact Lie group the {\em
semi-classical approximation is exact} (a discussion is given in
Ref.\cite{camporesi}). In the other words, the sum of contributions from
solutions of the classical equations of motion satisfies the wave equation
(\ref{ka}) with the initial condition,
       \begin{equation}%
\label{delta} 
\lim_{t\rightarrow 0}K_t(g)=\delta (g).
          \end{equation}
Here the $\delta$-function on $\sf G$ is defined as usual with the 
integration measure which is employed in Eq. (\ref{ka}). Note that every
separate term in Eq. (\ref{scap}) is a function on $\sf T^*$, the space
tangent to $\sf T$, and does not satisfy the desired boundary conditions
on  $\sf T$, but the series as a whole is indeed a function on the group.

For compact groups the spectrum of $-\Delta$ is positive 
semi-definite, i.e. $\lambda_{\bl}\ge 0$, and the series in (4) is 
convergent in the complex $t$-plane below the real axis. Because of the 
same reason, the heat transport equation has a stable solution.
For non-compact Lie groups, however, the spectrum is not positive, so
evidently the spectral expansion does not exist. On one hand, the series 
cannot be convergent even for complex $t$. On the other hand, the unitary 
representations are all infinite-dimensional, so $d_{\bl}$ would be 
infinite, as well as the volume of Eq. (\ref{volume}), while the
characters
are singular. The heat transport equation would have no stable solutions,
but one may still consider the wave equation,
determine the appropriate class of the wave functions and look for 
a valid representation of the evolution kernel. This is done 
in the present work.

\section{Green's function}
\label{green}
%\subsection{General arguments}
The operator equation (3) is equivalent to the Laplace transform for the
corresponding integral kernels, 
      \begin{equation}%(3.1)
\label{8}
K_t(h)=\frac{1}{2\pi i}\int_{\sf C}G_\lambda(h)
e^{-i\lambda t}d\lambda,\;\;\;
G_\lambda(g_1g_0^{-1})\equiv\langle g_1|\HG|g_0\rangle. 
      \end{equation}
Like $K_t(g)$, $G_\lambda(g)$ is a central function on the group.
It is Green's function for the (inhomogeneous) Helmholtz 
equation, which means that it solves the following problem,
    \begin{equation}   %(3.2)
\label{kf}
(\Delta+\lambda)\psi(g)=-f(g),\;\;\;
\psi(g_1)=\int_{\sf G}G_\lambda(g_1g_0^{-1}f(g_0)d\mu(g_0),
     \end{equation}
where the proper boundary conditions are taken into account.

The coordinates on the group manifold are introduced by the decomposition
$g=vhv^{-1}$, $\forall g\in{\sf G}$. The measure is factorized,
and the Laplacean is splitted respectively,
     \begin{eqnarray}   %(3.3)
\label{3.3}
\Delta=\Delta_{\sf T}+w^{-2}(h)\Delta_{\sf V}.
      \end{eqnarray}
where $\Delta_{\sf T}$ is the {\em radial} part and $\Delta_{\sf V}$ is
the {\em angular} part of the Laplacean; $w(h)$ is the Weyl function
on $\sf T$,
     \begin{equation}   %(12)
\label{ww}
w(h)\equiv\prod_{\balpha>0}\sin(\balpha\bvarphi/2),\;\;\;
h(\bvarphi)=\exp\left(i\varphi^j H_j\right)\in{\sf T}.
      \end{equation}
$H_j,\;j=1,...,r$ are the basis elements of the Cartan Subalgebra, 
$\varphi^j$ are the radial group parameters, which reside in the 
$r$-dimensional space tangent to {\sf T}.

As was shown by Berezin\cite{ber56}, the radial Laplacean may be 
reduced to the Euclidean form, as follows,
\begin{equation}  
\label{13}
\Delta_{\sf T}\equiv \frac{1}{\Lambda} 
\frac{1}{w^2}\frac{\partial}{\partial\bvarphi}
w^2\frac{\partial}{\partial\bvarphi}=
\frac{1}{\Lambda}\left[\frac{1}{w}\frac{\partial^2}{\partial\bvarphi^2}w+\rho^2\right].
      \end{equation}
(The constants $\Lambda$ and  $\rho^2$ depends on normalization, see Eq. \ref{rho}. In our 
convention,  $\rho^2/\Lambda=\frac{n}{24}$.)
Thus the desired Green's function can be reduced to that for the (pseudo)- 
Euclidean problem
     \begin{equation}  %14
\label{kd}
\frac{\partial^2 y}{\partial\bvarphi^2}+\epsilon y=-F,
      \end{equation}
where 
     \begin{equation}  
\label{ke}
y=w\psi,\;\;\;F= \Lambda wf,\;\;\;\epsilon=\rho^2+\Lambda \lambda.
      \end{equation}
The geometry is flat, yet the group structure manifests itself in the 
boundary conditions. In the two following sections, the boundary conditions 
are analyzed for compact and non-compact groups, the Green functions
are constructed explicitly and inserted in the integral representation Eq. \ref{8}
to produce the exact evolution operators.

\section{ Evolution operator on compact groups}
\label{grcom}
The boundary conditions are naively determined by requiring
the vanishing of $y$ on the hyper-surfaces of vanishing $w$.
The smallest domain enclosed by these hyper-surfaces in the 
r-dimensional root space is called the Weyl alcove\cite{bourbaki}.
(For example, the Weyl alcove of SU(3) is shown in fig. 2).
The Green function is found using the image method\cite{morse}
\begin{equation}
\label{gr1}
G^y(\bvarphi,\bvarphi_0)=\sum_{\sigma}\epsilon_\sigma
G^{r}(\sigma \bvarphi,\bvarphi_0).
\end{equation}
$G^{r}$ is the known Green function for Helmholtz Eq. in $r$-dimensional
flat infinite space $\IR^r$. The superscript $y$ is used to remind us that this is the Green
function for Eq. \ref{kd}.
The summation is over all reflections of the point $\bvarphi$ inside the Weyl
alcove, $\epsilon_{\sigma}= 1(-1)$ if the reflection is even (odd).

Since we are interested in the Green function itself,
the boundary conditions can be imposed on it directly. 
The boundary conditions, for which the Green function on the group manifold should 
account, are periodicity in the radial parameters, and  symmetry
under Weyl reflections (central functions are invariant under Weyl reflections,
which permute the eigenvalues of the group element). 
The multiplication by the Weyl function \ref{ke} 
which is antisymmetric under Weyl reflections imposes antisymmetry on  the 
Green function in the flat space, $G^y$,
\begin{equation}
\label{gy}
 G^{y}(\bvarphi,\bvarphi_0)=\sum_{\bm}\sum_{\sigma \in W}\epsilon_{\sigma}
G^{r}\left[\sigma(\bvarphi+2\pi\bm),\bvarphi_0\right]
\end{equation}
where $W$ is the Weyl group, $\sigma \in W$ is a Weyl reflection,
$\epsilon_\sigma=+1(-1)$ for $\sigma$ even (odd), and 
$\bm$ is the winding numbers vector
\be 
\label{mm}
\bm=\sum_{i=1}^r m_i \frac{2\bgamma_i}{\gamma_i^2},\;\;\;m_i=0,\pm1,\pm2,...
\ee 
$\bgamma_i,\;(i=1,...,r)$ are the simple roots.
The Lattice of images that is created by reflections of a point inside the Weyl
alcove of SU(3) is shown in fig. 3.
It is important to note (for future calculations) 
that there are two equivalent ways to perform the summation 
\begin{equation}
\sum_{\bm}\sum_{\sigma \in W}\epsilon_{\sigma}f\left[
\sigma(\bvarphi+2\pi\bm)\right]=
\sum_{\bm}\sum_{\sigma \in W}\epsilon_{\sigma}f\left[
\sigma\bvarphi+2\pi\bm\right].
\end{equation}
$G^y(\bvarphi,\bvarphi_0)$ is the Green function for Helmholtz equation 
in flat space with nontrivial boundary conditions, and not the resolvent that 
appears in the integral representation Eq. \ref{8}. The resolvent $G_{\lambda}(\bvarphi)$,
where $\bvarphi$ are the radial parameters that correspond to the group element $g_1g_0^{-1}$,
is obtained  by
taking into account the substitutions \ref{ke}, the different integration measures 
on the group manifold and the flat root space, and the fact that the radial parameters already
represent the `distance' between two points on the manifold so the point $\bvarphi_0$ should 
be set to zero (the procedure is explained in more details in appendix \ref{app7}). On the other
hand, it may be more convenient to continue the computation with $G^y$,
insert $G^y$ into the integral representation \ref{8} and  get the evolution operator
in flat $r$ dimensional space. The evolution operator on the group manifold is then 
obtained by the same procedure
\begin{equation}
\label{kt}
G_{\lambda}(\bvarphi)=\frac{\Lambda}{V_{{\sf G/T}}
w(\bvarphi)}\left[\frac{G^y(\bvarphi,\bvarphi_0)}{\Lambda^{r/2}2^{n-r}w(\bvarphi_0)}
\right|_{\bvarphi_0=0},\;\;\;\;
K_t(\bvarphi)=\frac{\Lambda}{V_{{\sf G/T}}
w(\bvarphi)}\left[\frac{K^y(\bvarphi,\bvarphi_0)}{\Lambda^{r/2} 2^{n-r}w(\bvarphi_0)}
\right|_{\bvarphi_0=0}
\end{equation}
$V_{\sf G/T}$ is the volume of the quotient space ${\sf G/T}$ (Eq. \ref{volgt}) in the compact case,
and a normalization constant in the non-compact case.

The integration \ref{8} can be performed in two alternative ways
leading to the known expressions \ref{spex},\ref{scap} for the evolution operator. 
Using the residue method, the spectral expansion is reconstructed.
Integrating the infinite sum \ref{gy} term by term we get an exact expression
for the evolution operator which is interpreted as the sum over classical paths.
Up till now, the sum over classical paths was built using the semi-classical approximation,
and its exact equivalence to the spectral expansion was proven by using multidimensional
Theta function theorems \cite{bellman,mlong}. The fact that this  expression
can be computed directly using no approximations is a new result that will be useful
in the non-compact case.

\subsection{Sum over classical paths}

We shall use the following integral representation for the
Green function for Helmholtz equation in flat $r$ dimensional 
infinite space
\begin{equation}
\label{grgr}
G_k^{r}(\bvarphi,\bvarphi_0)=\frac{1}{(2\pi)^r}\int_{\IR^r}
d^r \bp\frac{1}{p^2-k^2}\exp\left[i\bp\cdot(\bvarphi-\bvarphi_0)\right],
\end{equation}
$k^2=\rho^2+\Lambda\lambda$ in our case, and $\bp$ is an $r$-dimensional
vector. 
Inserting one term of the infinite sum \ref{gy} into the integral 
representation for the evolution operator and changing the order of integration yields
\begin{eqnarray}
K^{r}(\bvarphi,\bvarphi_0)&=&\frac{1}{2\pi i}
\int_{{\sf C}}e^{-i\lambda t}G^{r}_{\rho^2+
\Lambda \lambda}(\bvarphi,\bvarphi_0)d\lambda= \nonumber \\ &=&
\frac{1}{2\pi i}\frac{1}{(2\pi)^r}\int_{\IR^r}
d^r \bp\exp\left[i\bp\cdot(\bvarphi-\bvarphi_0)\right]
\int_{{\sf C}}e^{-i\lambda t}\frac{1}{p^2-\rho^2-\Lambda\lambda}d\lambda.
\end{eqnarray}
To  account for the singularities of the resolvent
which lie on the real axis in the complex
$\lambda$ plane, the integration contour {\sf C} passes infinitesimally
above the real axis, and it  closes underneath at infinity where the 
integrand vanishes (for $t>0$). Performing the integration in $\lambda$ by the residue method
we get
\begin{eqnarray}
K^{r}(\bvarphi,\bvarphi_0)&=&
\frac{1}{(2\pi)^r}\int_{\IR^r}d^r \bp\exp\left[i\bp\cdot(\bvarphi-\bvarphi_0)
-i(p^2-\rho^2) \frac{t}{\Lambda} \right]=\nonumber \\ &=&
\left(\frac{\Lambda}{4\pi i t}\right)^{r/2}
\frac{1}{\Lambda}e^{i\frac{\rho^2}{\Lambda}t}
\exp\left[\frac{i\Lambda}{4t}(\bvarphi-\bvarphi_0)^2\right].
\end{eqnarray}
Thus, the infinite sum (Eq. \ref{gy}) becomes
\begin{equation}
\label{ky}
K^{y}(\bvarphi,\bvarphi_0)=\left(\frac{\Lambda}{4\pi i t}\right)^{r/2}
\frac{1}{\Lambda}e^{i\frac{\rho^2}{\Lambda}t}
\sum_{\bm} \left\{\sum_{\sigma \in W}\epsilon_{\sigma}
\exp\left[\frac{i\Lambda}{4t}\left(\sigma(\bvarphi+
2\pi\bm)-\bvarphi_0\right)^2\right]\right\}.
\end{equation}
To get the exact evolution operator on the group manifold, $K^y$ is 
inserted into Eq. \ref{kt}. However, both the numerator and denominator
vanish when $\bvarphi_0=0$. The distance from the origin of all the Weyl reflections
of the point $\bvarphi+2\pi\bm$ is equal, and due to the factor
$\epsilon_{\sigma}$ the term in curled brackets in Eq. \ref{ky} vanishes. 
The Weyl function in the denominator is a product
of $p$ (the number of positive roots) sines which result in a pole
of order $p$ when $\bvarphi_0=0$. There are two 
ways (which are actually the same) to resolve the problem.
The first way is to set $\bvarphi_0=t\bzeta$ where $\bzeta$ is an {\em arbitrary}
vector in the root space, and let $t\rightarrow 0$. Then the usual L'Hopital rule
can be used
\be
\lim_{\bvarphi\rightarrow 0}\left[\frac{K^y(\bvarphi,\bvarphi_0)}{w(\bvarphi_0)}\right]
=\lim_{t\rightarrow 0}\frac{\frac{\partial^p}{\partial t^p}K^y(\bvarphi,t\bzeta)}{
\frac{\partial^p}{\partial t^p}w(t\bzeta)}
\ee
 This method is used in obtaining the celebrated Weyl dimension  formula
(see e.g. Ref. \cite{cahn}). It is important to stress that the actual choice of $\bzeta$ 
does not change the result since we are dealing with functions on the torus that possess
the special symmetry of the root space. A convenient choice is to take $\bzeta=\brho$
(convenient in the sense that it is easier to work with this choice analytically).

An alternative method is to to act directly on the functions in the numerator and the denominator
by any $p$-order differential operator on the torus. Again, the actual choice of the operator
is not important. We can differentiate $p$ times
along a specific direction, which is the same as the previous method,
or we can take an arbitrary combination of $p$ directional derivatives and to obtain the same result.
We chose to work with the $p$-order operator  ${\cal D}$
(see appendix \ref{app8}) that has some   special features
\begin{equation}
\label{op}
{\cal D}(\bvarphi)=\prod_{\balpha>0}\left[\frac{2\balpha_j}{\alpha_j^2}\cdot \sum_{i=1}^r
 \bw_i \frac{\partial}{\partial \varphi_{i}}\right],
\end{equation}
where $\bw_i$ are the fundamental weights, and $\varphi_i$ are the components of the vector 
$\bvarphi$ in the natural basis
\be
\label{nat}
\bvarphi=\sum_{i=1}^r \varphi_i \bgamma_i,
\ee
and $\bgamma_i$ are the primitive roots (see appendix \ref{app1}).
 When $\cal D$  operates on a function
with a definite symmetry under Weyl reflections (i.e. a function which is
either symmetric or anti-symmetric under Weyl reflections), it changes the symmetry.
Both the Weyl function and $K^y(\bvarphi,\bvarphi_0)$ are antisymmetric under Weyl
reflections, and therefore they vanish when on the hyper-surfaces of the Weyl
reflections, and in particular when $\bvarphi_0\rightarrow 0$. Acting
on them with $\cal D$ turn them into symmetric functions, and therefore different
from zero as $\bvarphi_0\rightarrow 0$.
Acting with ${\cal D}(\bvarphi_0)$ on the Weyl function and taking the limit $\bvarphi_0=0$
we get
\begin{equation}
\left.{\cal D}(\bvarphi_0)w(\bvarphi_0)\right|_{\bvarphi_0=0}=\frac{1}{2^p}N(W)\prod_{
\balpha>0}(\balpha \cdot \brho),
\end{equation}
where $N(W)$ is the order of the Weyl group. Acting on $K^y$ with ${\cal D}$
and setting $\bvarphi_0=0$ we get
\begin{eqnarray}
\left.{\cal D}(\bvarphi_0)K^{y}(\bvarphi,\bvarphi_0)\right|_{\bvarphi_0=0}&=&
\left(\frac{\Lambda}{4\pi i t}\right)^{r/2}
\frac{1}{\Lambda}e^{i\frac{\rho^2}{\Lambda}t}
\sum_{\bm}N(W)\left(\frac{-i\Lambda }{2t}\right)^{p}\times \nonumber \\ &\times& \prod_{
\balpha>0}\left[\balpha \cdot (\bvarphi+2\pi \bm)\right]
\exp\left[\frac{i\Lambda}{4t}\left(\bvarphi+
2\pi\bm\right)^2\right].
\end{eqnarray}
Inserting all the factors in the final expression for the evolution operator
Eq. \ref{kt} together with the volume of the quotient space $V_{{\sf G/T}}$
(Eq. \ref{volgt}),
we get
\begin{equation}
K_t(\bvarphi)=\left(\frac{1}{4\pi i t}\right)^{n/2}\sum_{\bm}
\left\{\prod_{\balpha>0}\frac{\balpha\cdot (\bvarphi+2\pi \bm)}{2\sin\frac{\balpha
\cdot\bvarphi}{2}}\right\}
\exp\left[i\frac{\Lambda}{4t}(\bvarphi+2\pi \bm)^2
+i\frac{\rho^2}{\Lambda}t\right].
\end{equation}
This is an exact expression for the evolution operator since it is 
calculated directly using no approximations, yet it can be interpreted as the sum
over classical paths (Eq. \ref{scap}). The first term in the exponent is 
the classical action, the pre-exponential factor 
(in curled brackets) is `the first quantum correction' (Van-Vleck determinant),
and the second term in the exponent is `the second quantum correction' which
is proportional to the scalar curvature on the manifold ($\rho^2/\Lambda=
n/24=R/6$).

\subsection{Spectral expansion}

We shall use again the integral representation for the Green function in flat $r$ dimensional
infinite space Eq. \ref{grgr}. However, in this case choosing the appropriate basis for the vector $\bp$
becomes important.
To simplify the computation, it is beneficial to use a basis
 which is dual to the natural basis of $\bvarphi$ (Eq. \ref{nat}), i.e. to use the fundamental
weights $\bw_j$ as the basis vectors for $\bp$: $\bp=\sum_{j=1}^r ip_j\bw_j$.
Thus, the scalar product in the exponential is $i\sum_jp_j(\varphi_j\gamma_j^2/2+2\pi m_j)$.
The integration measure is $d^r\bp=\left[{\rm det}(\bw_i \bw_j)\right]^{1/2}\prod_i dp_i$.
The  evolution operator is 
\begin{equation}
K^{y}(\bvarphi,\bvarphi_0)=\frac{\left[{\rm det}(\bw_i \bw_j)\right]^{1/2}
}{ i(2\pi)^{r+1}}
\int_{-\infty}^{\infty}d\lambda e^{-i\lambda t}\sum_{\bm}\sum_{\sigma\in W}
\epsilon_{\sigma} \int_{-\infty}^{\infty}\prod_i dp_i
\frac{\exp\left[i\bp\cdot\left(\sigma\bvarphi+2\pi \bm-\bvarphi_0\right)\right]}{
p^2-\rho^2-\Lambda\lambda}
\end{equation}
Changing the order of integration and performing the integral in $\lambda$
by the residue method one gets
\begin{eqnarray}
K^{y}(\bvarphi,\bvarphi_0)&=&\frac{\left[{\rm det}(\bw_i \bw_j)\right]^{1
/2}}{\Lambda(2\pi)^{r}}\sum_{\sigma\in W}
\epsilon_{\sigma}\int_{-\infty}^{\infty}\prod_i dp_i
\exp\left[i(p^2-\rho^2)t/\Lambda+i \bp \cdot \left(
\sigma \bvarphi-\bvarphi_{0}\right)\right]\times \nonumber \\ &\times&
\prod_{i=1}^r\left[
 \sum_{m_i=-\infty}^{\infty}\exp(2\pi i m_i p_i)\right]
\end{eqnarray}
The integration in $p_i$ can be easily performed by using the following
identity
\begin{equation}
\int_{-\infty}^{\infty}dp_i\sum_{m_i=-\infty}^{\infty}\exp(2\pi i m_i p_i)f(p_i)=
\sum_{n_i=-\infty}^{\infty}f(p_i=n_i)
\end{equation}
where $f$ is an arbitrary smooth function and $n_i$ are integers. 
Thus, the evolution operator becomes
\begin{equation}
K^{y}(\bvarphi,\bvarphi_0)=\frac{\left[{\rm det}(\bw_i \bw_j)\right]^{1/2}
}{\Lambda(2\pi)^{r}}\sum_{\bn}\sum_{\sigma\in W}
\epsilon_{\sigma}\exp\left[-i(n^2-\rho^2)t/\Lambda\right]
\exp\left[i\bn\cdot(\sigma\bvarphi-\bvarphi_0)\right].
\end{equation}
The vector $\bn$ is defined on the weights lattice 
$\bn=\sum_{i}n_i\bw_i$. The Weyl reflections permute the weights as well
as the roots, so it is possible to sum on the vectors $\bn$ in the Weyl
chamber (where the components $n_i$ are positive integers, see appendix \ref{app5}), and act
on each vector with the Weyl group to reproduce the rest of the weight lattice
\begin{equation}
K^{y}(\bvarphi,\bvarphi_0)=\frac{\left[{\rm det}(\bw_i \bw_j)\right]^{1/2}
}{\Lambda(2\pi)^{r}}\sum_{\bn>0}
e^{-i(n^2-\rho^2)t/\Lambda}\left\{\sum_{\sigma_1\in W}
\epsilon_{\sigma_1}\sum_{\sigma_2\in W}
\epsilon_{\sigma_2}
e^{i\sigma_1\bn\cdot(\sigma_2\bvarphi-\bvarphi_0)}\right\}.
\end{equation}
To obtain the evolution operator on the group manifold $K^y$ should be inserted
into Eq. \ref{kt}.
When $\bvarphi_0=0$ the term in curled brackets vanishes. Thus, we
have to act on it with the operator ${\cal D}$ (Eq. \ref{op}) before taking
the limit
\begin{equation}
\left.{\cal D}K^{y}(\bvarphi,\bvarphi_0)\right|_{\bvarphi_0=0}=
\frac{N(W)(-i)^p }{\Lambda(2\pi)^{r}}
\left[{\rm det}(\bw_i \bw_j)\right]^{1/2}\sum_{\bn>0}
e^{-i(n^2-\rho^2)t/\Lambda}\prod_{\balpha>0}
(\balpha\cdot \bn)\sum_{\sigma\in W}
\epsilon_{\sigma}e^{i(\bn\cdot \sigma\bvarphi)}.
\end{equation}
The final result is 
\begin{equation}
K_t(\bvarphi)=\frac{\left[{\rm det}(\bw_i \bw_j)\right]^{1
/2}}{V_{{\sf G/T}}(2\pi)^{r}\Lambda^{r/2}}\sum_{\bn>0}
\left[\prod_{\balpha>0}\frac{(\balpha\cdot\bn)}{(\balpha\cdot\brho)}\right]
\left[\frac{1}{(2i)^p w(\bvarphi)}\sum_{\sigma\in W}
\epsilon_{\sigma}\exp\left[i(\bn\cdot\sigma\bvarphi)\right]\right]
e^{-i(n^2-\rho^2)t/\Lambda}
\end{equation}
The factor in the exponent coincides with the known spectrum of the Casimir operator
($-\Delta$). Identifying  $\bn=\bl+\brho$ where $\bl$ is the highest weight
of the representation, $\lambda_{\bl}=n^2-\rho^2$. The two factors in square brackets
are the dimensionality and the character of each representation (compare
with Eqs. \ref{dim},\ref{char}). Since $\bn$ is defined over the weight lattice, and the
summation is limited to vectors with positive coefficients $n_i$, the summation is easily
translated to a summation over the highest weights $l_i=n_i-1$.
The numerical factor in front of the first summation is the inverse of
the group volume (see eqs. \ref{volgt},\ref{volt}).

This expression coincide with the known expression for the 
spectral expansion \ref{spex}.
 Thus, the Weyl formulas for the characters and dimensionalities of the UIR are 
obtained here by direct computation of the evolution operator.

\section{Evolution operator on non-compact group manifolds}
\label{grncom}
The method for obtaining the evolution operator using an integral representation,
described in section \ref{grcom} for compact groups,
works equally well for non-compact groups. The difference
between a compact group and its non-compact partners is that for the
compact group, the killing metrics (eq. \ref{a6})
is positive definite, while for the non-compact groups it is not (a method for obtaining all the noncompact 
groups having the same complex extension from a compact group
 is given in section \ref{sec31}). It is natural to expect that the indefinite
metric on the group manifold, and in the corresponding algebra space, induces
indefinite metric in the tangent space ${\sf T}^*$  where the radial parameters reside.
Therefore, the flat $r$-dimensional space of the radial parameters
is no longer Euclidean but pseudo-Euclidean $\IR^r\rightarrow\IR^{a,b}$,
where there are $a$ `space-like' radial parameters and $b$ `time-like'
radial parameters, and $a+b=r$. The vector $\bvarphi$ that is built
of the radial parameters can be rearranged in two sub-vectors
\[\bvarphi=(\bphi,i\btheta)=(\phi_1,...,\phi_a,i\theta_1,...,i\theta_b).\]
 It should be noted that the two subspaces
are orthogonal, so the natural coordinate system for the radial parameters \ref{nat},
where the radial parameters are defined along the primitive roots, 
 is not appropriate in this case.
 The maximal torus cease to be a closed torus and becomes an open torus
${\sf T}^r\rightarrow{\sf T}^{a,b}={\sf T}^a\otimes \IR^{b}$. The periodicity
of the group element in the radial parameters 
is altered due to the fact that $b$ of the radial parameters are no longer real
\begin{equation}
h(\bvarphi)=\exp\left[i\sum_{j=1}^r(\varphi_jH_j)\right]
=\exp\left[i\sum_{j=1}^a(\phi_jH_j)-\sum_{k=1}^b(\theta_k H_k)\right].
\label{Da}
\end{equation}
$H_j$ are the basis elements of the Cartan subalgebra.
  Thus, the  winding number vector in the non-compact case is altered with respect to the compact case (eq. \ref{mm})
$\bm\rightarrow \tilde{\bm}$. $\tilde{\bm}$ is obtained from $\bm$ by requiring that it vanishes in the
subspace $\IR^b$, and therefore it  is an $a$-dimensional vector instead of the $r$-dimensional
vector in the compact case
\[ \bvarphi+2\pi \tilde{\bm}=(\bphi+2\pi \tilde{\bm},\btheta).\]
Now we can proceed along the lines of the previous section. 
The first step is to find the 
Green function for Helmholtz equation in $r$-dimensional flat (pseudo-Euclidean)
space (eq. \ref{kd}) with nontrivial boundary conditions. 
The boundary conditions, that are imposed on the evolution operator, are that it should be symmetric 
under Weyl reflections,  that it should account for the periodicity {\em in the periodic 
radial parameters}, and that it should  decrease properly in the open domains.
Thus, we can use the appropriate Green function for Helmholtz equation in the flat 
infinite space $\IR^{a,b}$, and sum over the equivalent points, as we did in the compact case
\be
\label{grab}
G^y(\bvarphi,\bvarphi_0)=\sum_{\sigma \in W} \epsilon_\sigma 
\sum_{\tilde{\bm}} G^{a,b}\left[ \sigma (\bvarphi+2\pi \tilde{\bm}),\bvarphi_0\right].
\ee
This expression is inserted into the integral representation (Eq. \ref{8}).
 Because of the incompleteness
of the winding number vector, the integration method that reconstruct the spectral
expansion in the compact case is not applicable, 
which is to be expected since we know that the 
spectral expansion does not exist in the non-compact case. Yet, we can integrate 
term by term and reconstruct the `sum over classical paths'.

The Green function for Helmholtz equation in infinite 
pseudo-Euclidean  space that appears in Eq. \ref{grab}  has the general
form 
\begin{equation}
G_{\epsilon}^{a,b}(\bvarphi,\bvarphi_0)=\frac{1}{(2\pi)^{(a+b)}}\int_{\IR^{a,b}}
d^r \bp\frac{1}{p^2-\epsilon}\exp\left[i\bp\cdot(\bvarphi-\bvarphi_0)\right],\;\;\;\epsilon=
\rho^2+\Lambda\lambda,
\end{equation}
where the `momentum' vector $\bp$ has $a$ `space like' coordinates and $b$
`time-like' coordinates,
 \[ \bp=(\bq,i\bk)=(q_1,...,q_a,ik_1,...,ik_b).\]
 Inserting one term in the sum
\ref{grab} into the integral representation for the evolution operator
and changing the order of integration we get
\begin{equation}
K^{a,b}(\bvarphi,\bvarphi_0)=\frac{1}{2\pi i}\frac{1}{(2\pi)^{r}}\int_{\IR^{a,b}}
d^r \bp\exp\left[i\bp\cdot(\bvarphi-\bvarphi_0)\right]
\int_{{\sf C}}e^{-i\lambda t}\frac{1}{p^2-\rho^2-\Lambda\lambda}d\lambda.
\end{equation}
Unlike the compact case, $p^2$ is not positive definite, but the single pole at $\lambda=(p^2-\rho^2)/\Lambda$
is located on the real axis of the complex $\lambda$ plane, and the integral in $\lambda$
is solved by the residue method (the integration contour ${\sf C}$ is defined as in the compact
case).
Thus, we are left with the integral over $\IR^{a,b}$ which can be divided into 
two separate integrals over the subspaces $\IR^a,\IR^b$
\begin{eqnarray}
K^{a,b}(\bvarphi,\bvarphi_0)&=&\frac{1}{(2\pi)^{r}}\frac{1}{\Lambda}
e^{i\rho^2\frac{t}{\Lambda}}\int_{\IR^{a,b}}
d^r \bp\exp\left[i\bp\cdot(\bvarphi-\bvarphi_0)-itp^2/\Lambda\right]\nonumber \\&= &
\frac{1}{\Lambda}e^{i\rho^2\frac{t}{\Lambda}}\left[\frac{1}{(2\pi)^{a}}\int_{\IR^{a}}
d^a \bq\; e^{i\bq\cdot(\bphi-\bphi_0)-itq^2/\Lambda}\right]
\left[\frac{1}{(2\pi)^{b}}\int_{\IR^{b}}
d^b \bk\; e^{-i\bk\cdot(\btheta-\btheta_0)+itk^2/\Lambda}\right]\nonumber \\ &= &
\frac{1}{\Lambda}e^{i\rho^2\frac{t}{\Lambda}}\left[\left(\frac{\Lambda}{4\pi i t}\right)^{a/2}
\exp\left[\frac{i\Lambda}{4t}(\bphi-\bphi_0)^2\right]\right]
\left[\left(\frac{\Lambda}{4\pi i t}\right)^{b/2}
\exp\left[-\frac{i\Lambda}{4t}(\btheta-\btheta_0)^2\right]\right]=\nonumber \\ &= &
\left(\frac{\Lambda}{4\pi i t}\right)^{r/2}
\frac{1}{\Lambda}e^{i\frac{\rho^2}{\Lambda}t}
\exp\left[\frac{i\Lambda}{4t}\left(\bvarphi-\bvarphi_0\right)^2\right].
\end{eqnarray}
Integrating over Eq. \ref{grab} term by term, 
the evolution operator in the flat space is obtained
\be
K^{y}(\bvarphi,\bvarphi_0)=\left(\frac{\Lambda}{4\pi i t}\right)^{r/2}
\frac{1}{\Lambda}e^{i\frac{\rho^2}{\Lambda}t}
\sum_{\tilde{\bm}} \left\{\sum_{\sigma \in W}\epsilon_{\sigma}
\exp\left[\frac{i\Lambda}{4t}\left(\sigma(\bphi+
2\pi\tilde{\bm},i\btheta)-(\bphi_0,i\btheta_0)\right)^2\right]\right\}.
\ee
To get the evolution operator on the group manifold it should be inserted into 
Eq. \ref{kt}. Due to the anti-symmetry under Weyl reflections of both
the numerator and denominator, we have to reverse the symmetry when taking the limit $\bvarphi_0=0$
with the help of the operator  ${\cal D}$ (Eq. \ref{op}). The operator  is  unchanged  
except for the fact that  the parameters $\varphi_i$ are not real in general, and it should be
re-expressed in terms of the real parameters $\phi_i,\theta_i$. 
To produce the correct normalization of the evolution operator, the factor $V_{{\sf G/T}}$
should remain as in the compact case.
The final expression is 
 \begin{eqnarray}
\label{encom}
K_t(\bvarphi)&=&\left(\frac{1}{4\pi i t}\right)^{n/2}\sum_{\tilde{\bm}}
\left\{\prod_{\alpha>0}\frac{\balpha\cdot (\bvarphi+2\pi \tilde{\bm})}{2\sin\frac{\balpha
\cdot\bvarphi}{2}}\right\}
\exp\left[i\frac{\Lambda}{4t}\left(\bvarphi+2\pi \tilde{\bm}\right)^2
+i\frac{\rho^2}{\Lambda}t\right]\nonumber \\
&=&\left(\frac{1}{4\pi i t}\right)^{n/2}\sum_{\tilde{\bm}}
\left\{\prod_{\alpha>0}\frac{\balpha\cdot (\bphi+2\pi \tilde{\bm},i\btheta)}{2\sin\frac{\balpha
\cdot(\bphi,i\btheta)}{2}}\right\}
\exp\left[i\frac{\Lambda}{4t}\left((\bphi+2\pi \tilde{\bm})^2
-\btheta^2\right)+i\frac{\rho^2}{\Lambda}t\right].
\end{eqnarray}
 Once more we emphasize that
although this expression has a semi-classical interpretation, it is an exact 
expression that is obtained using no approximations. 

Up to this point, the discussion was limited to a specific configuration of the radial parameters.
However, when considering the evolution operator on non-compact groups a complication arises.
The manifold of a non-compact group is splitted, in most cases, into several domains, and 
in each domain the maximal torus  topology is different. Therefore,
a global spherical coordinate system on the manifold of non-compact groups
does not exist in these cases. There are several coordinate patches on the manifold, and they differ
by the number of radial parameters in which the group element is 
periodic, i.e. the number of the real radial parameters. 
Thus, the first step in finding the evolution operator on a non-compact group manifold is
 to determine the different coordinate patches
(we shall call them {\em evolution domains}) which differ 
by the decomposition of the radial parameters vector $\bvarphi$ into the sub-vectors 
$\bphi,i\btheta$.
In figs. 4,5 we can see for example the lattices of equivalent points 
(that have to be summed upon, see Eq. \ref{gr1})
 for SU(2,1) and SL(3,\IR), respectively.
Each of these non-compact groups has two evolution domains, which differs from each
other by the winding numbers vector $\tilde{\bm}$ (see the results section \ref{nunu}).
The different topology of the torus in each domain affects the winding numbers vector,
i.e. the periodicity in the radial group parameters, hence the lattices are different.
A full analysis of the different coordinate patches  of certain
`families' (e.g. SU(p,q), SO(p,q), etc.) of non-compact real groups
is given in section \ref{sec32}. 

The different periodicity in the radial group parameters affects  the boundary conditions
that are imposed on the evolution operator.
 Thus, the evolution operator   should be determined separately in each
 domain. The evolution of a state $\Psi(g)$
on the manifold, given an initial state $\Psi(g)|_{t=0}=
\psi(g)$,  is predicted by the following integral
\be
\label{psi}
 \Psi(g_t)=\sum_{a}\int_{D_{a}}K_{a}(g_tg_0^{-1})\psi(g_0)dg_0.
\ee
$D_a$ represent the domain to which the group element $g=g_tg_0^{-1}$ belongs,
and the index $a$ is the number of real radial parameters in which the group element is periodic.
The element $h \in {\sf T}$,  that corresponds to  $g=vhv^{-1}\in D_a$, has the general form 
that is given in Eq. \ref{Da}.

\section{Non-compact real groups and evolution domains analysis}

Three subjects, that are required to better understanding of the the structure of non-compact
real groups, are discussed in this section. The first subject is a short summary 
of the general method of finding all the non-compact groups having the same complex extension 
from a compact group.  
The second subject concerns the coordinate patches on the noncompact group manifold
when spherical coordinate system is used (evolution domains).
The third subject concerns the use of the evolution operator. 
The evolution operator depends only on the radial parameters of $g=g_tg_0^{-1}$.
A simple method for connecting the general group element  $g=g_tg_0^{-1}$
and the evolution domain to which it
 belongs is explained.

\subsection{Classification of non-compact real groups}
\label{sec31}
The method of finding all the real groups associated with a compact group
and having the same complex extension (and the same dimensionality)
is equivalent to 
 finding all the involutive automorphisms of the compact form.
An involutive automorphism  is a linear one-to-one
transformation of the compact group on itself 
conserving the Lie multiplication
\[ S[x,y]=[Sx,Sy] \;\;\;\;(x,y\in {\cal G}) \]
and fulfilling
\[ S^2=1.\]
Choosing a  basis in the algebra space, which is 
 composed of eigenvectors of the matrix $S$, multiplying
 the eigenvectors that correspond to eigenvalue $-1$ by
$i$ and leaving the rest of the eigenvectors unchanged, a non-compact
real group is obtained \cite{gant39a,gant39b,barut1}. 
An equivalent way of finding the real non-compact forms associated with a compact
Lie algebra is to look for the direct-sum decompositions 
of the compact Lie
Algebra\cite{hermann}. 
${\cal G}$ has a direct sum decomposition of the form ${\cal G}=K\oplus P$
if the following commutation relations hold:
\begin{equation}
\label{aaa}
 [K,K]\subset K,\;\;\; [K,P]\subset P, \;\;\; [P,P]\subset K.
\end{equation}
A non-compact real form is obtained by multiplying  the generators
that constitute $P$ by an $i$: $P\rightarrow iP$. 
The form of the commutation relations \ref{aaa} ensures 
that the structure constants remain real.
This method 
may be easier to grasp than the more general method given above,
however it is basis dependent and thus may not
give the full list of possible direct-sum decompositions and the corresponding
non-compact real forms. 

An important distinction should be made between inner and outer automorphisms
of the compact group (an inner automorphism is an 
isomorphic mapping of the group ${\sf G}$ into itself by a fixed element 
$x$ of the group: $z'=xzx^{-1}$, for all $z \in {\sf G}$. It induces an inner 
automorphism of the corresponding Lie algebra. All other automorphisms are outer
automorphisms). Inner automorphisms lead to real Lie groups with 
similar structures, while
the outer automorphisms lead in general to different real  
Lie groups. The existence of outer automorphisms is related to invariance
 of the corresponding Dynkin diagrams under transformations other than the 
identity transformation \cite{barut1,tits}. As we shall see in section \ref{secres},
 in the case of  non-compact groups that correspond to inner automorphisms
some of the evolution operators in the different domains coincide, and in 
particular there is always one evolution domain where the operator is identical
to that in the compact case. Yet, such coincidences do not exist in 
non-compact groups that correspond to outer automorphisms, although
they may occur when comparing  groups that both correspond 
to outer automorphisms. 
A non-compact group correspond to an inner (outer) automorphism if the determinant of 
$S$ is equal to $+1(-1)$, or if the number of generators  that belong to the subset
$P$ is even (odd).
The  algebras that have outer as well as inner 
automorphisms are $A_n$, $D_n$  and $E_6$ \cite{barut1,gant39b,tits}.

\subsection{Characteristic evolution domains}
\label{sec32}
As was mentioned in the previous section, a global spherical 
coordinate system does not exist in general for noncompact groups.
Thus, the group manifold is divided into several domains,
and the evolution operator is found separately in each domain.
Analyzing the eigenvalue system of the group element, which depends
solely on the radial parameters, shows the allowed values of the radial parameters
in the non-compact case, and each range of values corresponds to a different evolution domain.
In this section, an analysis of the evolution domains of several
families of groups is given. 

\subsubsection{Quasi-Unitary groups SU(p,q)}

The real group SU(p,q) (p+q=n) corresponds to an inner automorphism of the compact group SU(n).
It is obtained by leaving the generators of the subgroup SU(p)$\otimes$SU(q)$\otimes$U(1)
invariant (these generators constitute the maximal compact subalgebra of SU(p,q) denoted by $K$)
and multiplying the rest of the generators by an $i$.

A quasi unitary group element  $g$ (in the fundamental representation)
satisfies the relation
\begin{equation}
\label{yh}
g\eta g^{\dagger}=\eta
\end{equation}
where $\eta$ is a constant $n\times n$ matrix (in the compact case $\eta=I$)
with p eigenvalues equal to $+1$ and  q eigenvalues equal to $-1$.
An understanding of the structure of the evolution domains is achieved by contemplating the 
eigenvalue system of the group element $g$, which depends  on the $n-1$ radial parameters.
The characteristic polynomial of a unitary group element is
$P_{\lambda}(g)={\rm det}(\lambda I-g)=\prod_i(\lambda-\lambda_i)$.
The complex conjugate of the characteristic polynomial can be written in two alternative
forms
\begin{eqnarray}
P_{\lambda}^{*}(g)&=&\prod_i (\lambda-\lambda_i^*)\nonumber \\
&=&{\rm det}(\lambda I-g^{\dagger})={\rm det}(\lambda I-\eta^{-1} g^{-1}\eta )= 
{\rm det}(\lambda I-g^{-1})=\prod_j (\lambda-\lambda_j^{-1}),
\end{eqnarray}
which leads to the following relation
\begin{equation}
\label{za}
\lambda_i^*=\lambda_j^{-1},
\end{equation}
i.e.  for any eigenvalue $\lambda_i$ there must be another eigenvalue
$\lambda_j$ that satisfies  relation \ref{za}. 
In the compact case, where all the radial parameters are real,
this relation is satisfied for $i=j$.
For non-compact real groups, there are several possibilities. In general, there are
min(p,q)+1 evolution domains corresponding to $0,1,...,{\rm min(p,q)}$ imaginary 
radial parameters. The eigenvalues should be written in the following way:
\[ \exp[i(\varphi_1+\varphi_2)],\exp[i(\varphi_1-\varphi_2)],\exp[i(-\varphi_1+\varphi_3)],
\exp[i(-\varphi_1-\varphi_3)],...\]
 where $\varphi_1, \varphi_2, \varphi_3,...$ are the
radial parameters. It is clear that $\varphi_1$ must remain real, while $\varphi_2, \varphi_3$
can be either pure real or imaginary in the above example
(for an odd number of eigenvalues the last one must have a real radial parameter,
i.e. a pure imaginary argument). This choice of radial parameters
corresponds to an orthogonal coordinate system in ${\sf T}^*$.

\subsubsection{Real Unimodular groups SL(n,\IR)}

The real group SL(n,\IR) corresponds to an outer automorphism of the compact group
SU(n). The compact subgroup is SO(n), and the generators that remain unchanged (i.e. 
belong to $K$) are the generators of SO(n) in the vector representation (the spinor
representation of SU(n) is of the same dimension of the vector representation of
SO(n) and among its generators there are n generators which are
pure imaginary), and
the rest of the generators are multiplied by an $i$.
The fundamental group element is a n-dimensional real matrix whose
determinant is equal to one. The invariants of this group element (e.g
${\rm Tr}(g), {\rm Tr(g^2)}$, etc.) must
be real, and this fact  imposes conditions on the radial parameters
of the group. Thus, the eigenvalues of the group element must be either
complex conjugate to each other, or real:
\begin{equation}
 \lambda_i=\lambda_j^* 
\label{aab}
\end{equation}
The first domain is the domain in which all eigenvalues are real so  
the radial parameters are pure imaginary.
The rest of the domains correspond to taking pairs of eigenvalues to
be complex conjugate.
The maximal number of allowed domains is [n/2]+1.
It is important to note that since SL(n,\IR) (for n$>2$) correspond to an outer 
automorphism of SU(n) \cite{barut1}, 
while the different SU(p,q) correspond to 
inner ones, none of the domains of the radial parameters in SL(n,\IR)
coincide with the domains of SU(p,q).

\subsubsection{Quasi-Orthogonal groups SO(p,q), p+q=2n+1}

The real group SO(p,q) corresponds to an inner automorphism of the compact group
SO(2n+1). The maximal compact subgroup is SO(p)$\otimes$SO(q), and the rest of the generators
are multiplied by an $i$ to form the group SO(p,q) (all these groups correspond to the 
complex algebra $B_n$ which has no outer automorphisms). Thus, the number of non-compact
groups that can be formed out of SO(2n+1) equal the number of possible divisions of 2n+1
into p and q.

A pseudo-orthogonal group element (in the vector representation which is p+q dimensional) 
satisfies the relation 
\begin{equation}
g\eta g^{T}=\eta
\end{equation}
where $\eta$ is a constant matrix. In the  compact case $\eta$
is simply the unit matrix, so $g^{T}=g^{-1}$, and
the eigenvalues of $g$ are the same as those of $g^{-1}$. This means that for every eigenvalue
$\lambda_i$ of $g$, there exists another eigenvalue of $g$, $\lambda_j$, that
satisfies the relation 
\begin{equation}
\label{zm}
\lambda_i=\lambda_j^{-1}.
\end{equation}
 However, this is also true
in the pseudo-orthogonal case, since $g^{T}$ and $g^{-1}$ are connected by a similarity
transformation. Since there are $n$ radial parameters,
the eigenvalues (in the compact as well as noncompact case) 
are divided into pairs of $e^{i\varphi_i}$, $e^{-i\varphi_i}$ so the condition
$\lambda_i=\lambda_j^{-1}$ is satisfied by each pair. Since the number of eigenvalues is odd
(p+q=2n+1), the last  eigenvalue is simply $\lambda_{2n+1}=1$. The radial parameters
have a simple physical interpretation in this case.
SO(p,q) is the rotation group in p+q Minkowski space.
Each pair of eigenvalues corresponds to an
independent rotation plane (in the 2n+1 dimensional space there are $n$ independent
rotation planes). Therefore a pure rotation can be performed in 
a plane that is consisted of two `space-like' or two `time-like' axes, or
 an hyperbolic rotation can be performed in a plane which is consisted of one
 `time-like' axis and one `space-like' axis.
 Therefore, a radial parameter that appears in the exponential $\lambda_i=\exp(i\varphi_i)$
can be either  real or pure imaginary, 
depending on the type of the rotation. 
It is important to note that condition \ref{zm} holds for complex parameters also, however a complex
parameter can be treated as two real parameters, so the next pair of eigenvalues will 
also depend on the same two real parameters (the total number of real radial parameters is fixed).
 A rotation of the coordinate system of the radial parameters in the root space 
transforms the eigenvalue system to the desired form. 

There can be a maximum
of $d={\rm min}(p,q)$ independent hyperbolic rotations, 
and therefore there will be
$d+1$ domains on the group manifold, the first corresponds to pure rotations
only, the second corresponds to a hyperbolic rotation in one plane and pure
rotations in the rest of the independent planes, etc., while the last domain 
corresponds to the maximal number of hyperbolic rotations with $d$ imaginary
radial parameters. It should be noted that it does not matter in this case which of 
the radial parameters is imaginary and which is real, since the action of the Weyl group 
permutes the radial parameters. Since the evolution operator is symmetric under the action 
of the Weyl group, the information that is needed to determine the evolution domain is 
the total number of imaginary parameters.

\subsubsection{Quasi-Orthogonal groups SO(p,q), p+q=2n}

The real groups SO(p,q) correspond to  inner as well as outer
automorphisms of the compact group, and this fact complicates the determination of the 
evolution domains. Even p and q correspond to an inner automorphism, while odd p and q
correspond to an outer  automorphism. Condition \ref{zm} holds in this case, and we can
relate each pair of eigenvalues in the $2n$ dimensional  vector representation to a rotation
in an independent plane (there are $n$ independent rotation planes and $n$ radial parameters),
however this case is more complicated then the previous one. 
An intuitive understanding of the profound difference
 between the groups that correspond to inner and outer automorphisms of the compact group 
is gained when considering the types of possible rotations in p+q Minkowski
space. When p and q are even, we can divide separately the time like part and the 
space like part into independent rotation planes, so that all the $n$ radial parameters correspond 
to pure rotations. All the group elements of this kind belong to a domain where all 
the radial parameters are real. In this domain the evolution operator coincide with
the evolution operator in the compact case, and this kind of non-compact 
group correspond to an inner automorphism of the compact group.
On the other hand, when p and q are odd, the maximal number of
pure rotation planes is $n-1$, and at least one radial parameter 
in each domain must correspond to a hyperbolic rotation. 
None of the domains is similar to the compact case, and the group correspond to
an outer automorphism.

Another subtlety concerns the general subspace with metrics $++--$ in the p+q
flat vector space where the group acts.  There
are three possible evolution domains; 
the first correspond to two pure rotations
in the $++$ and $--$ planes (with e.v. $e^{i\phi_1},e^{-i\phi_1},e^{i\phi_2},
e^{-i\phi_2}$), the second correspond to two hyperbolic rotations
in the $+-$ planes (with e.v. $e^{\phi_1},e^{-\phi_1},e^{\phi_2},
e^{-\phi_2}$), but there is also the possibility of one real parameter and one 
imaginary parameters in the rotated coordinate system
(with e.v. $e^{i\phi_1+\phi_2},e^{-i\phi_1-\phi_2},e^{-i\phi_1+\phi_2},
e^{i\phi_1-\phi_2}$). Determination of the appropriate coordinate system
of the radial parameters in the root space can be made using the method that 
is described in section \ref{evol2}. From observation, 
none of the evolution domains of groups that correspond 
to outer automorphisms coincide with those of groups 
that correspond to inner automorphisms.
\label{evol1}

\subsubsection{Quasi-Unitary Symplectic groups USp(2p,2q)}

In order to leave the two bilinear forms invariant, the group element
of USp(2p,2q), where 2p+2q=2n, has to satisfy the following conditions
\begin{equation}
g^{\dagger}\eta g=\eta,\;\;\;\;g^{T}\zeta g=\zeta.
\end{equation}
$\eta$ is a diagonal constant matrix of the form
\[ \eta={\rm diag}(I_p,-I_q,I_p,-I_q), \]
(in the compact case $\eta=I_{2n}$), and $\zeta$ is of the `metric tensor'
for the symplectic bilinear form 
\[ \zeta=\left(\begin{array}{cc}0& I_n\\-I_n& 0 \end{array}\right). \]
The two conditions should be satisfied simultaneously, and in particular
the eigenvalues of the group element in the $2n$-dimensional fundamental
representation should satisfy the conditions
\begin{eqnarray}
g^{\dagger}=\eta g^{-1} \eta^{-1}:& &\;\;\;\;\lambda_i^*=\lambda_j^{-1}\\
g^{T}=\zeta g^{-1} \zeta^{-1}:& &\;\;\;\;\lambda_i=\lambda_k^{-1}.
\end{eqnarray}
In the compact case, the first condition is satisfied automatically, $\lambda_i^*=
\lambda_i^{-1}$. To satisfy the second condition, the eigenvalues are divided 
into pairs of the form $e^{ia},\;e^{-ia}$ (where $a$ is real).
 
In the non-compact cases, the radial parameters may assume complex values. 
The eigenvalues are divided into quartets 
of the form 
\[ e^{i(a+b)},\;\; e^{i(a-b)},\;\; e^{i(-a+b)},\;\; e^{i(-a-b)},\]
where either $a$ or $b$ may assume pure imaginary values, but not both.
For each quartet the two conditions are satisfied when one of the 
parameters is real and the other is imaginary.
 On the manifold of  USp(2p,2q) there are $|$p-q$|+1$ domains,
in the first domain all the radial parameters are real and divided into pairs
as in the compact case, in the second domain there is one quartet which includes
one real parameter and one imaginary parameter, and the rest of the parameters 
are real, etc.

\subsubsection{ Real Symplectic groups Sp(2n,\IR)}

The group element of Sp(2n,\IR) in the fundamental representation is a 
real matrix that satisfies the relation
\[ g^{T}\zeta g=\zeta. \]
The eigenvalues satisfy the conditions
\[ \lambda_i=\lambda_j^*,\;\;\;\;\lambda_i=\lambda_k^{-1}. \]
Dividing the eigenvalues into pairs 
$\lambda_i=e^{ia},\;\lambda_{j}=e^{-ia}$,
we see that the two conditions are satisfied by each pair when the radial parameter
is either pure real or pure imaginary. Therefore, there are $n+1$ domains on
the manifold of Sp(2n,\IR) that correspond to $0,1,...,n$ imaginary parameters.

\subsection{Evolution domains on the group manifold}
\label{evol2}
The connection between the general group element given in the `Cartesian-like' coordinate system
and the radial parameters is easier to  investigate in the algebra space.
The general algebra  element   in Cartesian coordinates is 
$$x=\xi_a X_a,\;\;\; a=1,..,n\;\;\;x\in{\cal G}$$
where $X_a$ are the group generators, $\xi_a$ are the group parameters and
$n$ is the group dimensionality. The algebra element invariants such as ${\rm Tr}x,{\rm Tr}x^2...$
depend only on  the radial parameters, for example
$${\rm Tr}X^2 ={\rm Tr}(\xi_a X_a)^2={\rm Tr}(\varphi_jH_j)^2\;\;\;j=1,..,r$$
where $H_j$ are the generators of the Cartan subalgebra.
The number of independent invariants equals the rank of the algebra $r$, 
thus an equation system for finding the radial parameters is obtained.
The independent invariants that constitute the equation system should be chosen with
care. 
The representation that should be used is
 the fundamental representation (the representation
that correspond to the {\em covering group}). This representation has in general more invariants
than are needed to obtain the radial parameters (since only $r$ of them are independent),
and choosing invariants of higher order would result in higher order equations
and more solutions to the equation system while there are only $N(W)$ equivalent
solutions for the radial parameters, i.e. there should be only $N(W)$ solutions to the 
equation system (
$N(W)$ is the order of the Weyl group).
For the groups that correspond to the  series $A_{r}$, the fundamental (spinor) representation
is $r+1$ dimensional, and the $r$ invariants that should be chosen are ${\rm Tr}x^2,...,
{\rm Tr}x^{r+1}$. In other cases the set of lowest order independent 
invariants should be chosen. It is important to note that
all the solutions correspond to the same group element, and they can be
 obtained from each other by the action of the Weyl group. 
This is the origin of the requirement that the evolution 
operator is symmetric under Weyl reflections, i.e. invariant under 
the action of the Weyl group.

Choosing an appropriate coordinate system for the radial parameters in the root
space according to  section \ref{sec32}, we can make sure that 
at least in one solution all the radial parameters are  pure real and
imaginary. Thus, to each domain on the group manifold $D_{a}$ ($a$ is
the number of the real parameters) there corresponds a certain range of the 
invariants, and we can relate directly the group element to the domain to which it belongs.

\section{Evolution operator on SU(2) and SU(1,1)}

The fundamental group element of SU(2) in the spherical coordinate system is 
\[ g=\cos \frac{\varphi}{2} I_2+i\sin \frac{\varphi}{2}(\vec{\sigma}\cdot\hat{n})=vh(\varphi)v^{-1}\]
where $\hat{n}$ is a unit vector in three dimensional Euclidean space, and $\vec{\sigma}=(\sigma_1,
\sigma_2,\sigma_3)$
is a vector composed of the three Pauli matrices (which are also the generators of SU(2)).
This group element correspond to a rotation at an angle $\varphi$ around the unit vector $\hat{n}$.
The eigenvalues of the group element depend only on the radial parameter, $\varphi$, 
\[ \lambda_1=e^{i\varphi/2},\;\;\;\lambda_2=e^{-i\varphi/2}.\]
Note that the trace of the group element in the  fundamental representation
  lies in the interval $[-2,2]$.

SU(1,1) is obtained from the generators of SU(2) by substituting $\sigma_1,\sigma_2\rightarrow
i\sigma_1,i\sigma_2$. The maximal compact subgroup is $U(1)$, and it is generated by $\sigma_3$.
Since the group is non-compact, the radial parameter $\varphi$ may assume non-real values.
However, condition \ref{za} implies that either $\lambda_1^{-1}=\lambda_1^*$,
 $\lambda_2^{-1}=\lambda_2^*$, or  $\lambda_1^{-1}=\lambda_2^*$.
In the first case, $\varphi$ is real as in the compact case, and in the second case 
$\varphi$ is pure imaginary. Thus, two evolution domains are obtained; $D_1$ 
where $\varphi=\phi$ is real, and $D_0$ where $\varphi=i\theta$, and $\theta$ is real.
The group elements that belong to $D_1$ are periodic in $\varphi$, while the group elements
that belong to $D_0$ are not.
The evolution operator has to be computed separately in each domain, since the
 boundary conditions that should be  satisfied are different.
The evolution of a state on SU(1,1) is given by the following integral
\be \Psi_t(g_t)=\int_{g_tg_0^{-1}\in D_1} K_{1}(\phi)\psi(g_0)d\mu(g_0) 
+\int_{g_tg_0^{-1}\in D_0} K_{0}(i\phi)
\psi(g_0)d\mu(g_0). \ee
Note that on $D_1$, $\phi$ is given explicitly by the relation
\[ \phi=\cos^{-1}\left[\frac{1}{2}{\rm Tr}(g_tg_0^{-1}) \right],\]
and on $D_0$, $\theta$ is given by the relation
\[ \theta=\cosh^{-1}\left[\frac{1}{2}{\rm Tr}(g_tg_0^{-1}) \right].\]

\subsection{ Evolution operator on  SU(2)}

The root space of SU(2) is one dimensional. Normalizing the root 
length to $|\alpha|=1$, the constants $\Lambda,\rho^2$ that appear
in the general expression for the radial Laplacean (Eq. \ref{13}) are $\Lambda=2$,
$\rho^2=1/4$, and the Weyl function is
 \[ w(\varphi)=\sin\frac{\varphi}{2}.\] 
We are interested in finding the Green function for the inhomogeneous equation
\be (\Delta_{\sf T}+\lambda)\psi(\varphi)=
\frac{1}{2}\left[\frac{1}{\sin\frac{\varphi}{2}}\frac{\partial^2}{\partial
 \varphi^2}\sin \frac{\varphi}{2} +  \frac{1}{4}  +2\lambda\right]\psi(\varphi)
=-f(\varphi)   \ee
with appropriate boundary conditions.
Substituting $y(\varphi)=w(\varphi)\psi(\varphi)$ we arrive at 
the one dimensional Helmholtz equation
 \be\left[  \frac{\partial^2 }{\partial \varphi^2} +k^2 \right]y(\varphi)  =-F(\varphi),\;\;\;
k^2= \frac{1}{4}+2\lambda, \;\;\;F=2\sin\frac{\varphi}{2} f \ee 
with boundary conditions  $y(\varphi=0)=y(\varphi=2\pi)=0 $.
The Green function can be found by the image method; two `conducting
walls' are placed at $\varphi=0,\;\varphi=2\pi$, and  a unit charge, that is  placed 
at $\varphi_0$ inside the domain $[0,2\pi]$, is reflected with respect
to the walls, and an infinite series of images is created.

The appropriate Green function is 
\be
\label{su2sum}
 G^y(\varphi,\varphi_0)= \sum_{n=-\infty}^{\infty}\left[ G^{1}(\varphi, 
 \varphi_0+4\pi n)-G^{1}(\varphi,-\varphi_0+4\pi n)\right] 
\ee
where $G^{1}(\varphi,\varphi_0)$ is the Green function for Helmholtz equation in  $\IR^1$
\be
 G^{1}(\varphi,\varphi_0)=\frac{i}{2k}e^{ik|\varphi-\varphi_0|}.
 \ee
The same expression for the Green function is obtain if we impose the
more general boundary conditions discussed in section \ref{grcom},
that the green function should account for the periodicity of the 
group elements in the radial parameter $\varphi$ and that it should be 
antisymmetric (so the  evolution operator would be symmetric) 
under Weyl reflections. Thus, we should sum over all the points that
differ from each other by a period of $4\pi$,
 and subtract all the points that are created by the action of the Weyl
group (i.e. reflection through the point $\varphi=0$) 
on the infinite series of periodic points.

Inserting $G^1$ into the infinite sum \ref{su2sum} and summing, the following
expression is obtained
\be
G^y(\varphi,\varphi_0)=\frac{\sin(k\varphi_<)\sin\left[k(2\pi-\varphi_>)\right]}{
k\sin(2k\pi)},\;\;\;\;\;\;\varphi_<={\rm min}(\varphi,\varphi_0),\;\;
\varphi_>={\rm max}(\varphi,\varphi_0)
\ee
The desired Green function for Helmholtz equation on the group manifold  is
obtained by inserting $G^y$ into Eq. \ref{kt}, substituting $\varphi_>=\varphi, \;\varphi_<=
\varphi_0$ and taking the limit $\varphi_0=0$. The invariant volume of the angular parameters 
$V_{{\sf G/T}}=8\pi$ (see Eq. \ref{volgt}). The final expression is 
\be G_{\lambda}(\varphi)=\frac{\sin k(2\pi-\varphi)}{8\sqrt{2}\pi
\sin 2k\pi \sin\frac{\varphi}{2}} \ee
This expression is inserted into the integral representation for 
the evolution operator (Eq. \ref{8}). The integration contour $\sf C$ can be contracted around
the poles of the integrand at $\lambda=\frac{1}{8}(n^2-1)$ where $n$ is an integer. Changing the 
 integration variable from $\lambda$ to $k$, the integration contour `opens' 
and we obtain the following integral representation for the evolution operator 
\be
\label{kkt}
 K_t(\varphi)=\frac{1}{2\pi i}\int_{-\infty+i\delta }^{\infty+i\delta}
e^{-\frac{i}{2}(k^2-\frac{1}{4}) t}
 \frac{\sin k(2\pi-\varphi)}{8\sqrt{2}\pi \sin 2k\pi
\sin\frac{\varphi}{2}} kdk. \ee
The integral can be performed using two  alternative methods,
which lead to the two expressions found by Schulman\cite{schulman}
for the evolution operator, the spectral expansion and the sum over classical paths. 

To obtain the known expression for the spectral
 expansion, the integration contour can be closed by going from infinity to minus infinity
below the real axis, and then dividing the result by a factor of two.
The integral is solved  by the residue method, and the spectral expansion is
\be K_t(\varphi)=\frac{1}{32\sqrt{2}\pi^2}\sum_{n=1}^{\infty}n\frac{\sin(n\varphi)}{\sin
 \varphi}e^{-\frac{i}{8}(n^2-1)t}. \ee
Returning to the integral \ref{kkt}, expanding  the denominator
\[ \frac{\sin k(2\pi-\varphi)}{ \sin (2k \pi)}=
(e^{ik(4\pi-\varphi)}-e^{ik\varphi})e^{-4ik\pi}
\sum_{m=0}^{\infty}e^{-4ik\pi m} \]
and integrating term by term, an alternative expression is obtained
\be K_t(\varphi)=\frac{1}{(4\pi i t)^{3/2}}\sum_{m=-\infty}^{\infty}
 \frac{\varphi+4\pi m}{2\sin \frac{\varphi}{2}}\exp\left[
 \frac{i}{2t}(\varphi+4\pi m)^2+\frac{it}{8}\right],\ee
and this expression coincide with the   sum over classical paths. Note, that this expression, 
that was formerly obtained only by using the semi-classical approximation (and was proved to be exact using $\Theta$-function theorems) is obtained here by a direct computation, 
and in fact it stems from the same integral 
representation that produces  the spectral expansion.

\subsection{Evolution operator for SU(1,1)}

\subsubsection{The evolution operator on $D_1$:}
The radial Laplacean in the first coordinate patch, where the radial parameter
$\varphi=\phi$ is real, is identical
to that of SU(2) since the constants that depend on the root system
do not change, and the Weyl function $w(\phi)=\sin \frac{\phi}{2}$ is the same 
\be \Delta_1=\Delta_{{\rm SU(2)}}.\ee
The boundary conditions are also unchanged since the Green
function $G_{\lambda}$ has to account for the unchanged periodicity of the group
elements in $\phi$, and to be symmetric under Weyl reflections.
Thus, the evolution operator on $D_1$ is identical
to that of SU(2)
\be
\label{ksu11}
 K_1(\phi)=K_{{\rm SU(2)}}(\phi)=
\frac{1}{(4\pi i t)^{3/2}}\sum_{m=-\infty}^{\infty}
 \frac{\phi+4\pi m}{2\sin\frac{ \phi}{2}}\exp \left[
 \frac{i}{2t}(\phi+4\pi m)^2+\frac{it}{8}\right]. \ee
The `sum over classical paths' form was chosen here out of the two equivalent 
expressions, since this expression can be `analytically continued' to the
expression that is obtained in  the second evolution domain.

\subsubsection{The evolution operator on $D_0$:}
In this domain $\varphi$ is pure imaginary. Setting $\varphi=i\theta$, the Weyl function becomes
 $w(\varphi=i\theta)=\sinh\frac{ \theta}{2}$. The inhomogeneous equation for which we have to find the 
Green function is
\be (\Delta_0+\lambda)\psi=\frac{1}{2}\left[ 
-\frac{1}{\sinh \frac{\theta}{2}}\frac{\partial^2}{\partial \theta^2}
\sinh \frac{\theta}{2} +\frac{1}{4}+2\lambda\right]\psi(\theta)=-f(\theta) \ee
Substituting $y(\theta)=w(\theta)\psi(\theta)$, we get the inhomogeneous Helmholtz equation
in $\IR^1$
\be \left[ \frac{\partial^2}{\partial\varphi^2}-(\frac{1}{4}+2\lambda)\right]y(\theta)=F(\theta).\ee
Since the periodicity of the group element in $\varphi=i\theta$ is ruined, we are left only with
the boundary condition that the Green function $G^y$ is antisymmetric under Weyl reflections. 
The Green function in the infinite space should be chosen with great care due to the minus 
sign before the factor $\frac{1}{4}+2\lambda$ to avoid singularities when performing the integration
 in the complex $\lambda$ plane (in the integral representation Eq. \ref{8}). Therefore,
we must distinguish between the two cases where $k^2=-(\frac{1}{4}+2\lambda)>0$ and $k^2<0$
\begin{eqnarray}
& &\left.  G^{1}_k(\theta,\theta_0)\right|_{ k^2=-(\frac{1}{4}+2\lambda)<0}
=\frac{i}{2\sqrt{\frac{1}{4}+2\lambda}}e^{-\sqrt{\frac{1}{4}+2\lambda}|\theta-\theta_0|},\nonumber \\ & &
\left. G^{1}_k(\theta,\theta_0)\right|_{k^2=-(\frac{1}{4}+2\lambda)>0}
=\frac{i}{2\sqrt{\frac{1}{4}+2\lambda}}e^{\sqrt{\frac{1}{4}+2\lambda}|\theta-\theta_0|}
\end{eqnarray}
Summing over the two Weyl reflections 
\be G^{y}(\theta,\theta_0)=G^{1}(\theta,\theta_0)-
G^{1}(\theta,-\theta_0)\ee 
and inserting $G^y$ into Eq. \ref{kt} we get the final 
expression for the resolvent $G_{\lambda}(\varphi)$
in this domain
\be \left.G_{\lambda}(\theta)\right|_{\frac{1}{4}+2\lambda>0}=
\frac{1}{8\sqrt{2}\pi\sinh \frac{\theta}{2}}e^{-\sqrt{\frac{1}{4}+2\lambda}\theta},\;\;\;\;
 \left.G_{\lambda}(\theta)\right|_{\frac{1}{4}+2\lambda<0}=\frac{1}{8\sqrt{2}\pi\sinh 
\frac{\theta}{2}}e^{\sqrt{\frac{1}{4}+2\lambda}\theta}\ee
Inserting into the resolvent into the integral representation Eq. \ref{8} we get
\be K_{0}(\theta)=\frac{1}{2\pi i}\frac{1}{8\sqrt{2}\pi\sinh\frac{ \theta}{2}}\int_{{\sf C}}\left[
e^{-i\lambda t}e^{\sqrt{\frac{1}{4}+2\lambda}\theta}\Theta(-\frac{1}{4}-2\lambda)+
 e^{-i\lambda t}e^{-\sqrt{\frac{1}{4}+2\lambda}\theta}\Theta(\frac{1}{4}+2\lambda) 
\right]d\lambda. \ee
$\Theta$ is the usual step function, $\Theta(x<0)=0,\;\Theta(x>0)=1$.
The second term does not contain neither cuts nor poles, so it does not contribute to the integral.
There is a cut in the first term, and contracting the integration contour around the cut  
the final result is obtained. 

The exact evolution operator in the domain $D_0$ is
\be K_{0}(\varphi=i\theta)=\frac{1}{(4\pi it)^{3/2}}
\frac{\theta}{2\sinh\frac{\theta}{2}} \exp\left[-\frac{i}{2t}\theta^2+\frac{it}{8}\right]. \ee
Together with $K_1(\varphi)$ (Eq. \ref{ksu11}) 
in the evolution domain $D_1$ where $\varphi$ is real ,
the evolution operator on the entire group manifold is found.

\section{Results}
\label{secres}
Finding the evolution operator on a non-compact group according to section \ref{grncom}
involves the following steps. First, the evolution domains have to be 
established according to section \ref{sec32}. Then the appropriate coordinate system 
for the radial parameters, where the parameters are either real or imaginary is 
determined in each domain. The next step is to determine the winding number 
vector $\tilde{\bm}$, which is inserted in the expression for the evolution operator
 in each domain (eq. \ref{encom}). In this section the method is demonstrated on the real
groups associated with the four simple  algebras, $A_2, B_2$ $A_3$ and $C_3$.

We shall use the following notations:
{\sf G} is the noncompact group under consideration, {\sf K} its maximal compact subgroup,
$K_j$ the generators of {\sf K}, $iP_j$ the generators of the coset space
${\sf G}/{\sf K}$,
$D_a$ the evolution domain that correspond to $a$ real radial parameters and $r-a$
imaginary parameters. The imaginary parameters $\varphi_j$ are given in terms of the real parameters
$\theta_j$, $\varphi_j=i\theta_j$.
The rotation matrices $L_{ab}$ in the 4-dim. spinor representation of SO(5) and SO(6)$\sim$
SU(4) are given in  Appendix \ref{app9}, and the basis for $C_3$ is given in  Appendix \ref{app10}.

\subsection{Groups associated with $A_2$}

\label{nunu}

{\bf Compact group:} SU(3)\\
{\bf Generators:} Gell-Mann matrices\cite{greiner}
 $\lambda_1,...,\lambda_8$ in 3-dim. spinor representation\\
{\bf Primitive roots (in $\IR^2$)}:
\[ \bgamma_1=\hat{x},\;\;\;\;\;\bgamma_2=-\frac{1}{2}\hat{x}+\frac{\sqrt{3}}{2}\hat{y}\]
{\bf Radial parameters vector}
\[ \bvarphi=\varphi_1\bgamma_1+\varphi_2\bgamma_2=\phi_1\hat{x}+\phi_2\hat{y},
\;\;\;\;\;\;\phi_1=\varphi_1-\frac{\varphi_2}{2},\;\;\;\;\;\phi_2=\frac{\sqrt{3}}{2}\varphi_2\]
{\bf Winding number vector:}
\[\bm=2 m_1\bgamma_1+2m_2\bgamma_2=(2m_1-m_2)\hat{x}+\sqrt{3}m_2\hat{y}\]
{\bf eigenvalues of the group element in the spinor representation}
\[e^{\frac{i}{2}(\phi_1+\phi_2/\sqrt{3})},\;\;e^{\frac{i}{2}(-\phi_1+\phi_2/\sqrt{3})},
\;\;e^{-\frac{i}{\sqrt{3}}\phi_2}\]
{\bf Noncompact groups:}\\
\begin{tabular}{||l||l||l||}\hline \hline
{\sf G} & SU(2,1) & SL(3,\IR) \\ \hline \hline
{\sf K} & SU(2)$\otimes$U(1) & SO(3)\\ \hline
$K_j$ & $\lambda_1,\lambda_2,\lambda_3,\lambda_8 $& $\lambda_2,\lambda_5,\lambda_7 $\\
$iP_j$& $i\lambda_4,i\lambda_5,i\lambda_6,i\lambda_7$& $i\lambda_1,i\lambda_3,i\lambda_4,
i\lambda_6,i\lambda_8 $\\ \hline
 $D_a$   &$D_2$: \hspace{0.5cm} $\bvarphi=\phi_1\hat{x}+\phi_2\hat{y}$
 &$D_1$:\hspace{0.5cm}
$\bvarphi=\phi_1\hat{x}+i\theta_2\hat{y}$\\
    &\hspace{1cm} $\tilde{\bm}=(2m_1-m_2)\hat{x}+\sqrt{3}m_2\hat{y}$&
\hspace{1cm} $\tilde{\bm}=2m_1\hat{x}$\\
\cline{2-3}
    & $D_1$:\hspace{0.5cm}  $\bvarphi=i\theta_1\hat{x}+\phi_2\hat{y}$ &$D_0$:\hspace{0.5cm}
 $\bvarphi=i\theta_1\hat{x}+i\theta_2\hat{y}$\\
    & \hspace{1cm}$\tilde{\bm}=2\sqrt{3}m\hat{y}$&\hspace{1cm} $\tilde{\bm}=0$\\ \hline
\end{tabular}

\subsection{Groups associated with $B_2$}
{\bf Compact group:} SO(5)\\
{\bf Generators:} 10 rotation matrices $L_{ab},\;\; a,b=1,...,5$  in the 4-dim. spinor representation (see appendix \ref{app9})\\
{\bf Primitive roots (in $\IR^2$)}:
\[ \bgamma_1=\hat{x}-\hat{y},\;\;\;\;\;\bgamma_2=\hat{y}\]
{\bf Radial parameters vector:}
\[ \bvarphi=\varphi_1\bgamma_1+\varphi_2\bgamma_2=\phi_1\hat{x}+\phi_2\hat{y},
\;\;\;\;\;\;\varphi_1=\phi_1,\;\;\;\;\varphi_2=\phi_1+\phi_2\]
{\bf Winding number vector:}
\[\bm= m_1\bgamma_1+2m_2\bgamma_2=m_1\hat{x}+(2m_2-m_1)\hat{y}\]
{\bf eigenvalues of the group element in the 4-dim. spinor representation}
\[e^{\frac{i}{2}(\phi_1+\phi_2)},\;\;e^{\frac{i}{2}(-\phi_1+\phi_2)},
\;\;e^{\frac{i}{2}(\phi_1-\phi_2)},\;\;e^{\frac{i}{2}(-\phi_1-\phi_2)}\]
{\bf eigenvalues of the group element in the 5-dim. vector representation}
\[e^{i\phi_1},\;\;e^{-i\phi_1},\;\;e^{i\phi_2},\;\;e^{-i\phi_2},1\]
{\bf Noncompact groups}\\ \\
\begin{tabular}{||l||l||l||}\hline \hline
{\sf G} & SO(4,1) & SO(3,2) \\ \hline \hline
{\sf K} & SO(4) & SO(3)$\otimes$SO(2)\\ \hline
$K_j$ & $L_{ab},\;\;a,b=1,...,4 $& $L_{ab},L_{45}$\\
$iP_j$& $iL_{a5},\;\;a=1,...,4$& $iL_{a4},\;iL_{a5},\;\; a,b=1,2,3$\\ \hline
 $D_a$    &$D_2$: \hspace{0.5cm} $\bvarphi=\phi_1\hat{x}+\phi_2\hat{y}$
 &$D_2$:\hspace{0.5cm}
$\bvarphi=\phi_1\hat{x}+\phi_2\hat{y}$\\
    &\hspace{1cm} $\tilde{\bm}=m_1\hat{x}+(2m_2-m_1)\hat{y}$&
\hspace{1cm} $\tilde{\bm}=m_1\hat{x}+(2m_2-m_1)\hat{y}$\\
\cline{2-3}
    & $D_1$:\hspace{0.5cm}  $\bvarphi=\phi_1\hat{x}+i\theta_2\hat{y}$ &$D_1$:\hspace{0.5cm}
 $\bvarphi=\phi_1\hat{x}+i\theta_2\hat{y}$\\
    & \hspace{1cm}$\tilde{\bm}=2m\hat{x}$&\hspace{1cm} $\tilde{\bm}=2m\hat{x}$\\
\cline{3-3}
& &$D_0$:\hspace{0.5cm} $\bvarphi=i\theta_1\hat{x}+i\theta_2\hat{y}$\\
& &\hspace{1cm} $\tilde{\bm}=0$ \\ \hline \hline
\end{tabular}

\subsection{Groups associated with $A_3\sim D_3$}
{\bf Compact group:} SU(4)\\
{\bf Generators:} Two alternative bases in the 4-dim. spinor representation\\
\begin{itemize}
\item Gell-Mann type\cite{greiner} $4\times 4$ matrices $\lambda_1,...,\lambda_{15}$
\item 15 rotation matrices $L_{ab},\;\; a,b=1,...,6$ in spinor 4-dim. representation
(see appendix \ref{app9})
\end{itemize}
{\bf Primitive roots (in $\IR^3$)}:
\[ \bgamma_1=\hat{x},\;\;\;\;\;\bgamma_2=-\frac{1}{2}\hat{x}+\frac{1}{\sqrt{2}}\hat{y}
-\frac{1}{2}\hat{z},\;\;\;\;\;\bgamma_3=\hat{z}\]
{\bf Radial parameters vector}
\[ \bvarphi=\varphi_1\bgamma_1+\varphi_2\bgamma_2+\varphi_2\bgamma_2
=\phi_1\hat{x}+\phi_2\hat{y}+\phi_3\hat{z},\]
\[ \varphi_1=\phi_1+\frac{\phi_2}{\sqrt{2}},\;\;\phi_2=\sqrt{2}\phi_2
,\;\;\varphi_3=\phi_3+\frac{\phi_2}{\sqrt{2}}\]
{\bf Winding number vector:}
\[\bm=2 m_1\bgamma_1+2m_2\bgamma_2+2m_3\bgamma_3
=(2m_1-m_2)\hat{x}+\sqrt{2}m_2\hat{y}+(2m_3-m_2)\hat{z}\]
{\bf eigenvalues of the group element in the spinor representation}
\[e^{\frac{i}{2}(\phi_1+\phi_2/\sqrt{2})},e^{\frac{i}{2}(-\phi_1+\phi_2/\sqrt{2})},
e^{\frac{i}{2}(\phi_3-\phi_2/\sqrt{2})},e^{\frac{i}{2}(-\phi_3-\phi_2/\sqrt{2})}\]
{\bf eigenvalues of the group element in the 6-dim. vector  representation}
\[e^{\frac{i}{\sqrt{2}}(\phi_1+\phi_3)},e^{-\frac{i}{\sqrt{2}}(\phi_1+\phi_3)},e^{\frac{i}{\sqrt{2}}(\phi_1-\phi_3)},
e^{-\frac{i}{\sqrt{2}}(\phi_1-\phi_3)},e^{i\phi_2},e^{-i\phi_2},\]
{\bf Noncompact groups}\\ \\
The non-compact groups that correspond to inner automorphisms of SU(4) are SU(3,1) and
SU(2,2). \\ \\
\begin{tabular}{||l||l||l||}\hline \hline
{\sf G} & SU(3,1)$\sim$SO$^*$(6) & SU(2,2)$\sim$SO(4,2) \\ \hline \hline
{\sf K} &SU(3)$\otimes$U(1) & SU(2)$\otimes$SU(2)$\otimes$U(1)\\ \hline
$K_j$ &$\lambda_1,...,\lambda_8,\;\lambda_{15} $&$ \lambda_a,\;a=1,2,3,8,13,14,15$\\
$iP_j$& $i\lambda_9,...,i\lambda_{14}$& $i\lambda_b,\;b=4,5,6,7,9,10,11,12$\\ \hline
 $D_a$  &$D_3$: \hspace{0.5cm} $\bvarphi=\phi_1\hat{x}+\phi_2\hat{y}+\phi_3\hat{z}$
 &$D_3$:\hspace{0.5cm}
$\bvarphi=\phi_1\hat{x}+\phi_2\hat{y}+\phi_3\hat{z}$\\ &
  %  \hspace{1cm} 
$\tilde{\bm}=(2m_1-m_2)\hat{x}+\sqrt{2}m_2\hat{y}+(2m_3-m_2)\hat{z}$ &
%\hspace{1cm} 
$\tilde{\bm}=(2m_1-m_2)\hat{x}+\sqrt{2}m_2\hat{y}+(2m_3-m_2)\hat{z}$\\
\cline{2-3}
    & $D_2$:\hspace{0.5cm}  $\bvarphi=i\theta_1\hat{x}+\phi_2\hat{y}+\phi_3\hat{z}$ &$D_2$:\hspace{0.5cm}
 $\bvarphi=i\theta_1\hat{x}+\phi_2\hat{y}+\phi_3\hat{z}$\\
    & \hspace{1cm}$\tilde{\bm}=2\sqrt{2}m_1\hat{y}+(2m_3-2m_1)\hat{z}$&\hspace{1cm}
 $\tilde{\bm}=2\sqrt{2}m_1\hat{y}+(2m_3-2m_1)\hat{z}$\\
\cline{3-3}
& &$D_1$:\hspace{0.5cm} $\bvarphi=i\theta_1\hat{x}+\phi_2\hat{y}+i\theta_3\hat{z}$\\
& &\hspace{1cm} $\tilde{\bm}=2\sqrt{2}m\hat{y}$ \\ \hline \hline
\end{tabular}\\ \\
The noncompact groups that correspond to outer automorphisms of SU(4) are SL(4,\IR)
and Q(2)$\sim$SO(5,1). Q(2) is a group that acts in 2-dim. quaternionic space\cite{barut1,chevalley}.\\
\begin{tabular}{||l||l||l||}\hline \hline
{\sf G} & SO(3,3)$\sim$SL(4,\IR) & SO(5,1)$\sim$Q(2) \\ \hline \hline
{\sf K} &SO(3)$\otimes$SO(3) & SO(5)\\ \hline
$K_j$ &$L_{12},L_{23},L_{13},L_{45},L_{46},L_{56} $&$ L_{ab},\;a,b=1,...,5$\\
$iP_j$& $iL_{a4},iL_{a5},iL_{a6},\;a=1,2,3$& $iL_{a6},\;a=1,...,5$\\ \hline
 $D_a$    &$D_2$: \hspace{0.5cm} $\bvarphi=\phi_1\hat{x}+i\theta_2\hat{y}+\phi_3\hat{z}$
 &$D_2$:\hspace{0.5cm}
$\bvarphi=\phi_1\hat{x}+i\theta_2\hat{y}+\phi_3\hat{z}$\\
    &\hspace{1cm} $\tilde{\bm}=2m_1\hat{x}+2m_3\hat{z}$&
\hspace{1cm} $\tilde{\bm}=2m_1\hat{x}+2m_3\hat{z}$\\
\cline{2-2}
    & $D_1$:\hspace{0.5cm}  $\bvarphi=i\theta_1\hat{x}+i\theta_2\hat{y}+\phi_3\hat{z}$ &\\
    &\hspace{1cm}
 $\tilde{\bm}=2m_3\hat{z}$ &\\
\cline{2-2}
& $D_0$:\hspace{0.5cm} $\bvarphi=i\theta_1\hat{x}+i\theta_2\hat{y}+i\theta_3\hat{z}$& \\
&\hspace{1cm} $\tilde{\bm}=0$ & \\ \hline \hline
\end{tabular}\\ \\

\subsection{Groups associated with $C_3$}
{\bf Compact group:} USp(6)\\
{\bf Generators:} $X_1,...,X_{21}$ (see Appendix \ref{app10}\\
{\bf Primitive roots (in $\IR^3$)}:
\[ \bgamma_1=\hat{x},\;\;\;\;\;\bgamma_2=-\frac{1}{2}\hat{x}+\frac{1}{2}\hat{y}
-\frac{1}{\sqrt{2}}\hat{z},\;\;\;\;\bgamma_3=\sqrt{2}\hat{z} \]
{\bf Radial parameters vector}
\[ \bvarphi=\varphi_1\bgamma_1+\varphi_2\bgamma_2+\bvarphi_3\bgamma_3
=\phi_1\hat{x}+\phi_2\hat{y}+\phi_3\hat{z} \]
{\bf Winding number vector:}
\[\bm=2 m_1\bgamma_1+2m_2\bgamma_2+m_3\bgamma_3=(2m_1-m_2)\hat{x}+m_2\hat{y}+
\sqrt{2}(m_3-m_2)\hat{z}\]
{\bf Eigenvalues of the group element in the fundamental  representation}
\[e^{\frac{i}{2}(\phi_1+\phi_2)},\;\;e^{\frac{i}{2}(-\phi_1+\phi_2)},\;\;
e^{\frac{i}{2}(\phi_1-\phi_2)},\;\;e^{\frac{i}{2}(-\phi_1-\phi_2)},\;\;
e^{\frac{i}{\sqrt{2}}\phi_3},\;\;e^{-\frac{i}{\sqrt{2}}\phi_3}\]
{\bf Noncompact groups:}\\
\begin{tabular}{||l||l||l||}\hline \hline
{\sf G} & USp(4,2) & Sp(6,\IR) \\ \hline \hline
{\sf K} & USp(4)$\otimes$USp(2) & SU(3)$\otimes$U(1)\\ \hline
$K_j$ &$X_1,X_2,X_3,X_4,X_{5},X_{10},X_{11},X_{12},X_{13},$ &$X_{2n+1},\;\;n=2,3,...,10$\\
 &$X_{16}, X_{17},X_{20},X_{21}$ &\\
$iP_j$&$iX_6,iX_7,iX_8,iX_9,iX_{14},iX_{15},iX_{18},iX_{19}$
 & $iX_1,iX_3,iX_{2n}\;\;n=1,...,10$ \\ \hline
 $D_a$   &$D_3$: \hspace{0.5cm} $\bvarphi=\phi_1\hat{x}+\phi_2\hat{y}+\phi_3\hat{z}$
 &$D_3$:\hspace{0.5cm}
$\bvarphi=\phi_1\hat{x}+\phi_2\hat{y}+\phi_3\hat{z}$\\
    &\hspace{1cm} $\tilde{\bm}=(2m_1-m_2)\hat{x}+m_2\hat{y}+$&
\hspace{1cm} $\tilde{\bm}=(2m_1-m_2)\hat{x}+m_2\hat{y}+$\\
& \hspace{2cm}$+\sqrt{2}(m_3-m_2)\hat{z}$& \hspace{2cm}$+\sqrt{2}(m_3-m_2)\hat{z}$\\
\cline{2-3}
    & $D_2$:\hspace{0.5cm}  $\bvarphi=i\theta_1\hat{x}+\phi_2\hat{y}+\phi_3\hat{z}$
 &$D_2$:\hspace{0.5cm}
 $\bvarphi=\phi_1\hat{x}+\phi_2\hat{y}+i\theta_3\hat{z}$\\
    & \hspace{1cm}$\tilde{\bm}=2m_1\hat{y}+\sqrt{2}(m_3-2m_1)\hat{z}$
&\hspace{1cm} $\tilde{\bm}=(2m_1-m_2)\hat{x}+m_2\hat{y}$\\ \cline{3-3}
& &$D_1$:\hspace{0.5cm}
 $\bvarphi=i\theta_1\hat{x}+i\theta_2\hat{y}+\phi_3\hat{z}$\\
& &\hspace{1cm} $\tilde{\bm}=\sqrt{2}m_3\hat{z}$ \\ \cline{3-3}
& &$D_0$:\hspace{0.5cm}
 $\bvarphi=i\theta_1\hat{x}+i\theta_2\hat{y}+i\theta_3\hat{z}$\\
& &\hspace{1cm} $\tilde{\bm}=0$ \\ \hline \hline 
\end{tabular}

\section{Conclusion}
We have shown that using the  integral representation (eq. \ref{8}) for the evolution operator
is a powerful method, that produces  exact evolution operators of free motion on the manifolds of both
compact and non-compact groups.

For  compact groups, the two complimentary representations
for the evolution operator, the spectral expansion and the sum over classical paths,
 are reproduced from the same integral representation 
by using two different integration methods. 

For non-compact groups, this method enables us to find the exact evolution operator.
A general expression for the evolution operator is given in eq.  \ref{encom}.
This expression depends on the root system of the specific group, on the radial parameters vector
and on the winding numbers vector of the radial parameters around the maximal torus.

A complication, that arise in the non-compact case,
 is that the maximal torus topology is not unique on the 
entire group manifold. This is very different from the compact case, where the maximal torus
is a closed torus, and the winding number vector is determined solely by the primitive roots
(eq. \ref{mm}). In the non-compact case, the torus becomes an open torus, 
and the manifold is splitted into several domains which differ from each other by the maximal torus topology.
The radial parameters, which reside in a space tangent to the torus, are no longer real,
however a coordinate system where the radial parameters are either pure real or pure imaginary 
can always be found. 
Thus, the general expression for the evolution operator 
is different in each domain, and using it
requires the knowledge of the winding numbers vector in the domain.

The various domains, that correspond to each non-compact group, are derived from the group type
(quasi-unitary, quasi-orthogonal, etc.).
In particular, the maximal torus depends only on the eigenvalue system of the group element
(since we are dealing with matrix groups, the notion of eigenvalues   usually corresponds to
the fundamental representation). The allowed values of the  radial parameters are derived from
the conditions that are imposed on the eigenvalues according to the group type. This analysis
was carried out is section \ref{sec32}
for most of the simple groups that are generated from the classical algebras $A_n,\;
B_n,\;C_n$ and $D_n$. 

The method was demonstrated on SU(2) and SU(1,1) by a straightforward computation of the evolution operators.
For the larger groups, that are generated by the algebras $A_2,\;B_2,\;A_3$ and $C_3$, 
only the final results are given. For each group, an appropriate coordinate system for the radial parameters
was chosen, and the different domains were found.
The  radial parameters vector and winding number vector
were written explicitly for each domain. These vectors can be inserted into  expression \ref{encom}
to produce the exact expression for the evolution operator in the specific domain.

The expression for the evolution operator on non-compact groups can be interpreted as a sum over classical paths.
The different evolution domains correspond to different classes of geodetics. The group manifold is open
in some dimensions, and compact in the others. Thus, the winding number vector depends on the `direction'
of the classical trajectory. 

\section*{ Acknowledgments }
The support to the research from the Technion V. P. R. Fund is gratefully
acknowledged.

\section*{Appendix}
\appendix
\section{Some notations}
We shall use the following notations: $r$ is the rank of the Algebra;
$n$ is the dimensionality; $p=(n-r)/2$ is the number of positive roots.
There is an ordered system of positive roots $\balpha_{\nu},\;\nu=1,...,p$;
 among them are simple roots (a basis in the root space) 
$\bgamma_j,\; j=1,...,r$, and the highest root $\balpha_1=\sum_{j=1}^r
a_j\bgamma_j$ where $a_j$ are positive integers.
The scale factor $\Lambda$ and the vector $\brho$ 
are involved in sum important formulae. They are expressed in 
terms of the roots
\begin{equation}
\label{rho}
\brho=\frac{1}{2}\sum_{\balpha>0}\balpha,\;\;\;\;\;
\Lambda=2r^{-1}\sum_{\balpha>0}\balpha^2;
\end{equation}
the sums are over all positive roots. The Cartan matrix has elements
\begin{equation}
M_{jk}=2\bgamma_j\cdot \bgamma_k/\gamma_j^2
\end{equation}
that are integers: $0,\pm1,\pm2,-3$.
The fundamental weights satisfy the  following relation with the simple roots
\begin{equation}
\bgamma_i \bw_j=\gamma_i^2/2 \delta_{ij}.
\end{equation}
The unitary representations of compact groups are represented by the highest
weight $\bl=\sum_i l_i \bw_i$, $l_i$ are integers. 
The eigenvalues of the  Laplacean (Casimir operator)
as well as the dimensionality of the representations are given in terms of 
highest weight
\begin{equation}
\label{dim}
\lambda_{\bl}=\frac{1}{\Lambda}\left(n^2-\rho^2\right),\;\;\;\;
d_{\bl}=\prod_{\balpha>0}\frac{\balpha\cdot \bn}{\balpha\cdot \brho},
\end{equation}
where ${\bf n}=\bl+\brho$. 

\section{ Lie groups: summary}
\subsection{ The Lie group and its Lie algebra}
\label{app1}
Let $\sf G$ be a Lie group and $\cal G$ its corresponding Lie algebra,
     \begin{equation}   %(a1)
{\cal G}\rightarrow{\sf G}: g(x)=\exp x,\;\;\;x\in{\cal G},g\in{\sf G}.      
\end{equation} 
 A basis $\{e_a\}$, where $1\leq a\leq n$, is introduced in 
$\cal G$, so that
     \begin{equation}   %(a2)
x=\xi^ae_a,\;\;\;[e_a,e_b]=C^c_{ab}e_c,
      \end{equation}
where $C^c_{ab}$ are the structure constants, and 
$[\cdot,\cdot]$ stands for the Lie product in  $\cal G$. 
It is assumed that  $\cal G$ is real, i.e. there exists a basis where
the structure constants are real.
The adjoint matrix representation of $\sf G$ acting 
in $\cal G$ is defined by
     \begin{equation}  %a3
\label{ap1}
g^{-1}\exp(\zeta^ae_a)g=\exp[\zeta^aA^b_a(g)e_b],
      \end{equation}
where the $n\times n$ matrix ${\bf A}(g)$ with elements $A^b_a(g)$ is 
given by the exponential of the structure constants,
     \begin{equation}   %(a4)
{\bf A}(g)=\exp{\bf X},\;\;\;{\bf X}\equiv\xi^a{\bf E}_a,\;\;\;
({\bf E}_a)^c_b\equiv C^c_{ba}=-C^c_{ab}.      
\end{equation} 
(In this section, bold-face capitals are used for matrices in the adjoint
representation).
The commutators of matrices ${\bf E}_a$ provide with the basis for the 
regular representation of $\cal G$, as follows from the Jacobi identity,
     \begin{equation}   %(a5)
[{\bf E}_a,{\bf E}_b]=C^c_{ab}{\bf E}_c.
      \end{equation}
The scalar product is defined in $\cal G$ by means of the
adjoint representation,
     \begin{equation}   %(a6)
\label{a6}
(x,y)\equiv -{\rm Tr}({\bf XY})=\xi^a\eta^b\Gamma_{ab},\;\;\;
\Gamma_{ab}\equiv-{\rm Tr}({\bf E}_a{\bf E}_b)=C^c_{ad}C^d_{cb}.
      \end{equation}
The Cartan -- Killing matrix $\Gamma$ may be reduced by a proper linear
transformation to a diagonal form $\Gamma_{ab}=\Lambda \eta_{ab}$, 
where $\Lambda$ is a scaling factor and $\eta_{ab}$ has 
$n_+$ eigen-values $+1$ and $n_-$ eigen-values $-1$. It is assumed
that $\cal G$ is (semi-)simple, so $\Gamma$ is non-degenerate, and
$n_++n_-=n$. If $n_+=n$, i.e. $\Gamma$ is positive definite, the real
group $\sf G$ is compact. In general, $n_+$ is the dimensionality of the
maximal compact subgroup $\sf G_+\subset G$.

 \subsection{ The Cartan -- Maurer form and the Lie derivatives.}
The (left-invariant) Cartan -- Maurer one-form on $\cal G$ is defined by
     \begin{equation}   %(a7)
\delta x\equiv g(x)^{-1}dg(x)\equiv\delta\xi^ae_a=d\xi^bB^a_b(x)e_a,
      \end{equation}
where the matrix ${\bf B}(x)$ is expressed in terms of the adjoint group 
representation,
      \begin{equation}   %(a8)
\label{a8}
{\bf B}(x)=\int_0^1{\bf A}(g_\tau)d\tau,\;\;\;g_\tau\equiv\exp(\tau x).
      \end{equation}
The Killing field $\nabla_a(x)$, dual to the form 
$\delta\xi^a$, is defined by
     \begin{equation}   %(a9)
\delta\xi^a\nabla_a(x)\equiv d\xi^a\partial_a:\;\;\;
\nabla_a(x)=L^b_a(x)\partial_b,
      \end{equation}
where $\partial_a\equiv\partial/\partial\xi^a$,
${\bf L}(x)=\left[{\bf B}(x)\right]^{-1}$.
The Lie derivatives are introduced in terms of the Killing fields; namely
for $z=\zeta^ae_a$ and $\forall x,z\in{\cal G}$, 
     \begin{equation}   %(a10)
\pounds_z(x)\equiv\zeta^a\nabla_a(x)=\zeta^aL_a^b(x)\partial/\partial\xi^b.\;\;\;
      \end{equation}
By virtue of the Cartan -- Maurer equation, at any given $x\in{\cal G}$, the 
commutators of the Lie derivatives provide with a representation of $\cal G$, 
     \begin{equation}   %(a11)
d\delta x+\delta x\wedge\delta x=0\rightarrow \;\;\;
[\pounds_y(x),\pounds_z(x)]=\pounds_{[y,z]}(x).
      \end{equation}
Writing that in the components one has, respectively,
     \begin{eqnarray}   %(a12)
\partial_aB^c_b-\partial_bB^c_a+B^{a'}_aB^{b'}_bC^c_{a'b'}=0, \nonumber\\  
\left[\nabla_a(x),\nabla_b(x)\right]
=C_{ab}^c\nabla_c(x); \;\;\;
L^{a'}_a\partial_{a'}L^c_b-L^{b'}_b\partial_{b'}L^c_a=
C^{c'}_{ab}L^c_{c'}.
      \end{eqnarray}
Note that as follows from the definition of the Lie derivative,
     \begin{equation}   %(a13)
\label{a13}
\nabla_a(x){\bf A}(g)={\bf E}_a{\bf A}(g).
      \end{equation}

 \subsection{ Group as a Riemannian manifold.}
The Riemannian metrics in $\sf G$ is induced naturally by the
(pseudo-) Euclidean metrics in its tangent linear space $\cal G$,
as given by Eq. (\ref{a6}). The invariant length element is defined by
        \begin{equation}%(a14)
dx^2\equiv (\delta x,\delta x)=d\xi^a d\xi^b\gamma_{ab}(x),
        \end{equation}
where $\gamma_{ab}(x)=B_a^c(x)B_b^d(x)\Gamma_{cd}$. 
The Riemannian structure is invariant under general coordinate
transformations, and under the shifts of the group
elements, $g\rightarrow gg_1$, in particular. The corresponding
invariant measure on the group manifold is
        \begin{equation}%(a15)
d\mu(g)=|\gamma(\xi)|^{1/2}d^n\xi,
         \end{equation}
where $|\gamma(\xi)|
\equiv\det(\gamma_{ab})=\left[\det{\bf B}(x)\right]^2\det(\Gamma)$.
This measure is proportional to Weyl's 
invariant measure on the group. For compact
groups, the total invariant volume exists, and the explicit expression
was given in Ref.\cite{minv} (see also appendix \ref{app6}).

 The invariant Laplace operator $\Delta$ is defined as usual since the
metric is non-degenerate. It can be also considered as a realization of
the invariant Casimir operator in the universal 
enveloping algebra of $\cal G$, represented by the Killing fields,
        \begin{equation}%(a16)
\Delta\equiv\frac{1}{|\gamma|}
\frac{\partial}{\partial\xi^a}|\gamma|\gamma^{ab}
\frac{\partial}{\partial\xi^b}=\Gamma^{ab}\nabla_a\nabla_b,
         \end{equation}
The operator $\Delta$ is elliptic for compact groups and hyperbolic for
(real) non-compact groups.
(Higher-order invariant differential operators may be constructed 
by means of $\Gamma^{ab}$, structure constants and $\nabla$, 
see also appendix
\ref{app8}.)

\subsection{ Group as a fiber bundle.}
Let ${\sf G}_1$ be a subgroup of $\sf G$. Introduce a basis in the
subalgebra ${\cal G}_1$, $\{e_j\}$ and a basis in
${\cal Z\equiv G\setminus G}_1$, $\{ e_{\alpha} \}$ where 
$1\leq j\leq n_1$, $1\leq\alpha\leq(n-n_1)$. If ${\sf G}_1$ is
semi-simple, the basis can be chosen to satisfy
$(e_j,e_\alpha)=0$. Now the basic relations are
     \begin{eqnarray}   %(a17)
\left[e_j,e_k\right]=C^l_{jk}e_l,\;\;\;
\left[e_j,e_\alpha\right]=C^\beta_{j\alpha}e_\beta,\nonumber\\
\left[e_\alpha,e_\beta\right]=C^j_{\alpha\beta}e_j+
C^\gamma_{\alpha\beta}e_\gamma.
      \end{eqnarray}
The resulting regular representation of ${\cal G}_1$ has a block
structure, since $({\bf E}_j)^k_\alpha=0=({\bf E}_j)^\alpha_k$. 
The same is true also for the adjoint group representation: 
${\bf A}(h)=\exp(\sum_{j=1}^{n_1}{\bf E}_j\eta^j)$, 
$\forall h\in{\sf G}_1$. 

The group elements can be decomposed as follows,
     \begin{equation}   %(a18)
g(x)=v(z)h(y)v(z)^{-1},
      \end{equation}
where $h\in{\sf G}_1$, $y\in{\cal H}\subset{\cal G}$, and $x(z)$ is a
representative of the equivalence class $v(z)\sim v(z)h_1$, 
$\forall h_1\in{\sf G}_1$. 
The Cartan -- Maurer one-form and the metric form are decomposed
respectively,
     \begin{eqnarray}   %(a19-a20)
\delta x=v\left(\delta y+h^{-1}\delta z-\delta z\right)v^{-1},\\
\label{a20}
(\delta x,\delta x)=(\delta y,\delta y)+
2\left[(\delta z,\delta z)-(\delta z, h\delta zh^{-1})\right]\nonumber\\
\equiv(\delta y,\delta y)+2\delta\zeta^\alpha\delta\zeta^\beta
\tilde{\Gamma}_{\alpha\beta}(y),
      \end{eqnarray}
where $\tilde{\Gamma}_{\alpha\beta}(y)=
\Gamma_{\alpha\beta}-A^{\alpha'}_\alpha(h)\Gamma_{\alpha'\beta}$. 
The first term is the metric induced in ${\sf G}_1$, 
while the metric induced in $\cal Z$ depends on $h$.

\subsection{Radial coordinates}
The case of a particular interest is ${\cal G}_1={\cal H}$ -- the Cartan
subalgebra, ${\sf G}_1={\sf H}$ -- the maximal Abelian subgroup,
and $n_1=r$. Now the metric induced in $\sf H$ is flat.
The $h$-dependence of the measure on the group is given in terms of the
adjoint representation, given by a central function which may be presented
in a manifestly invariant form,
     \begin{eqnarray}   %(a21)
\label{a21}
d\mu(g)=W^2(h)d\mu(h)d\mu(v),\\
W^2(h)={\rm det}'\left[{\bf I-A}(h)\right]=
\lim_{\omega\rightarrow 1}
\left(\frac{{\rm det}\left[\omega{\bf I-A}(g)\right]
}{(\omega-1)^r}\right).\nonumber
     \end{eqnarray}   
Here ${\rm det}'$ is the determinant for the block corresponding to 
$\cal Z$, and we note that, $\forall g$, the matrix ${\bf A}(g)$ has at
least $r$ eigen-values which are equal to $1$, corresponding to the
subgroup $\sf H$. Other eigenvalues of  ${\bf A}(g)$ are denoted by
$\lambda_\alpha=\exp\left[\pm(\balpha\cdot\bvarphi)\right]$, where $\balpha$
are positive roots of $\cal G$, i.e. real $r$-dimensional vectors, and
$\bvarphi$ is a vector in the dual space. For compact groups,  ${\bf A}(g)$
is unitary, $|\lambda_\alpha|=1$, so $\bvarphi$ is a real vector, so that
\begin{equation}%(a22)
W(h)=\prod_\alpha \left[2\sin\frac{(\balpha\cdot\bvarphi)}{2}\right].
\end{equation}
Note that W(h) is proportional to the Weyl function $w(h)$ (Eq. \ref{ww}).
For non-compact real groups, some of the roots $(\balpha\cdot\bvarphi)$
are complex conjugate pairs.

\subsection{Invariant volumes of compact groups}

The invariant volume of a compact group is\cite{minv} 
\begin{equation}
\label{volg}
V_{{\sf G}}=\int_{{\sf G}}d\mu(g)=\frac{\Lambda^{n/2}
(2\pi)^{p+r}\left[{\rm det}(M_{jk})\right]^{1/2}}{\prod_{\bgamma}(\gamma^2/2)^{1/2}
\prod_{\balpha>0} (\balpha\cdot \brho)}.
\end{equation}
The invariant volume is a product of the
 volume of the maximal torus {\sf T}
and the volume of the coset space {\sf G/T},  $V_{{\sf G}}=
V_{{\sf T}}V_{{\sf G/T}}$, and each factor can be computed separately
\begin{equation}
\label{volt}
V_{{\sf T}}=\Lambda^{r/2}\int_{{\sf T}}2^{n-r}\left[w(\bvarphi)\right]^2 d\bvarphi=
\Lambda^{r/2}\frac{(2\pi)^{r}\left[{\rm det}(M_{jk})\right]^{1/2}}{\prod(\gamma^2/2)^{1/2}}=
\frac{\Lambda^{r/2}(2\pi)^{r}}{\left[{\rm det}(\bw_i \bw_j)\right]^{1/2}},
\end{equation}
where $d\bvarphi=\prod d\varphi$ is the integration measure in the flat space, and
\begin{equation}
\label{volgt}
V_{{\sf G/T}}=\frac{(2\pi)^p  \Lambda^{p}}{\prod_{\balpha>0}(\balpha\cdot\brho)}.
\end{equation}
Note that all these volumes are invariant under a change of the root normalization.
\label{app6}

\subsection{Weyl group, Weyl chamber, Weyl alcove}
\label{app5}
The Weyl group $W$ is the group of permutations of the root system.
The elements of the Weyl group are called Weyl reflections. Its action 
also permutes the diagonal elements of $\varphi_j H_j$, where $H_j$ are the basis elements 
of the Cartan subalgebra and $\varphi_j$ are the radial parameters..
To avoid this sort of ambiguity,  the values of the 
radial parameters are restricted to the Weyl chamber by the condition
$\bgamma\cdot \bvarphi\geq 0$. The root space  in which $\bvarphi$ reside
is divided by the hyper-planes $\balpha \cdot\bvarphi= 0$ into $N(W)$ regions
congruent to the Weyl chamber ($N(W)$ is the order of the Weyl group).
The Weyl transformations permute these regions.

Any function on a compact group is periodical in the radial coordinates
\begin{equation}
f(\bvarphi+2\pi \bm)=f(\bvarphi), \;\;\;\bm=\sum_{j=1}^{r}m_j \hat{\bgamma}_j,
\;\;\;\hat{\bgamma}_j=2 \bgamma_j/\gamma_j^2
\end{equation}
where $m_j$ are integers, so the maximal torus for the group reside inside
a larger torus defined by 
\begin{equation}
\bvarphi=\sum_{j=1}^{r} \varphi_j \hat{\bgamma}_j,\;\;\;\;
-\pi<\varphi_j\leq \pi.
\end{equation}
The weyl reflection hyper-surfaces $\balpha\cdot \bvarphi= 0$ divide the torus
into $N(W)$ regions. 
The region enclosed by the hyper-surfaces
$\bgamma\cdot \bvarphi\geq 0$, $\balpha^1 \cdot \bvarphi\leq 2\pi$, where $\balpha^1$
is the highest root, is called the Weyl alcove.
The Weyl alcove coincide with the maximal torus. 

The characters of the unitary irreducible representations, represented by the highest
weight $\bl$, are given by the Weyl formula
\begin{equation}
\label{char}
\chi_l(\bvarphi)=\frac{1}{(2i)^p w(\bvarphi)}
\sum_{\sigma\in W}\epsilon_{\sigma}\exp\left[i(\sigma \bn,\bvarphi)\right]
\end{equation}
where the summation is over Weyl reflections, $\epsilon_{\sigma}=+1(-1)$
for even (odd) reflection.

\section{Invariant operators and the symmetry operator \protect${\cal D}$}
\label{app8}
An invariant operator is built according to the following theorem (Berezin \cite{ber56}):\\
{\em Let $P(\varphi_1,...,\varphi_r)$ be any polynomial on the Cartan subalgebra $H$,
invariant with respect to the Weyl group $W$. Consider the differential operator
$P(\partial/\partial \varphi^1,...,\partial/\partial \varphi^r)$ which is 
obtained by formal substitution of the operators $\partial/\partial \varphi^i$ in place of
 the coordinates $\varphi_i$ in the polynomial $P(\varphi_1,...,\varphi_r)$.
The operator 
\begin{equation}
\label{p}
\tilde{\Delta}(P)=\frac{1}{w(\bvarphi)}\left[P\left(\frac{\partial}{\partial \varphi^1},...,
\frac{\partial}{\partial \varphi^r}\right)\right]w(\bvarphi)
\end{equation}
is the radial part of some Laplace operator on the group.}\\
The converse of this theorem is also true. The coordinates $\varphi_i$ and $\varphi^i$ are dual
coordinate systems (if $\bvarphi=\varphi_i\bgamma_i$ they are 
connected by the Cartan matrix $\varphi^i=M_{ij}\varphi_j$).

We shall use the operator
\begin{equation}
{\cal D}(\bvarphi)=\prod_{\balpha>0}\left[\frac{2\balpha_j}{\alpha_j^2}\sum_{i=1}^r
 \bw_i \frac{\partial}{\partial \varphi_{i}}\right],
\end{equation}
which is consisted of a product of directional  derivatives along the positive roots.
This operator is called an intertwining operator since it intertwines the radial $\delta$-function
on the group manifold with the  $\delta$-function on the torus $\sf T$ (see Ref. \cite{camporesi}).
Substituting the square of the operator, ${\cal D}^2$, for the polynomial
$P$ in Eq. \ref{p},  an invariant operator on the group is obtained.
When acting on the character of an UIR with ${\cal D}^2$, its eigenvalue
is proportional to the square of the representation's dimensionality
\begin{eqnarray}
\frac{1}{w(\bvarphi)}{\cal D}^2 w(\bvarphi) \chi_l(\bvarphi)&=&\frac{1}{(2i)^p w(\bvarphi)}
{\cal D}^2 \sum_{\sigma\in W}\epsilon_{\sigma}\exp\left[i(\sigma \bn\cdot\bvarphi)\right]\nonumber \\ &=&
 \left[\prod_{\balpha>0}i(\bn \cdot \balpha)\right]^2\chi_l(\bvarphi)=
\left[\prod_{\balpha>0}i(\brho \cdot \balpha)\right]^2 d_{l}^2 \chi_l(\bvarphi)
\end{eqnarray}
The operator ${\cal D}$ is not an invariant operator on the group manifold,
but it corresponds to the one-dimensional antisymmetric representation of the 
Weyl group. Therefore it  is used to change the symmetry of radial functions
under Weyl reflections. In particular
\begin{eqnarray}
 \frac{1}{w(\bvarphi)}{\cal D} w(\bvarphi) \chi_l(\bvarphi)&=&\frac{1}{(2i)^p w(\bvarphi)}
{\cal D} \sum_{\sigma\in W}\epsilon_{\sigma}\exp\left[i(\sigma \bn\cdot\bvarphi)\right]=
\nonumber \\ & &\prod_{\balpha>0}i(\bn \cdot \balpha)
 \sum_{\sigma\in W}\exp\left[i(\sigma \bn\cdot\bvarphi)\right]=
\prod_{\balpha>0}i(\brho \cdot \balpha) d_{l} \sum_{\sigma\in W}
 \exp\left[i(\sigma \bn\cdot\bvarphi)\right].
\end{eqnarray}
Taking the limit $\bvarphi=0$ reproduces the celebrated Weyl dimensions formula 
(up to known factors).

Another interesting feature is that when working on the Weyl function itself,
the operator can be used to determine the order of the Weyl group
\begin{equation}
\frac{2^p}{(\balpha \cdot \brho)}\left.{\cal D}w(\bvarphi_0)\right|_{\bvarphi_0=0}
=N(W).
\end{equation}

\section{The connection between Green's function on {\sf T}  and on {\sf G}}
\label{app7}
The resolvent for Helmholtz equation in flat $r$ dimensional space  $G^y(\bvarphi,\bvarphi_0)$
 is used to find the state $y(\bvarphi)$ that is described by Eq. \ref{kd} 
\begin{equation}
\label{reso}
y(\bvarphi)=\int G^y_{\lambda}(\bvarphi, \bvarphi_0) F(\bvarphi_0)d\bvarphi_0.
\end{equation}
$d\bvarphi=\prod d\varphi$ is the integration measure in the flat space.
Since our original equation  for the state $\psi(\bvarphi)$ is
\be 
\frac{1}{\Lambda}\left[\frac{1}{w}\frac{\partial^2}{\partial\bvarphi^2}w+\rho^2+\Lambda\lambda\right]
\psi(\bvarphi)=-f(\bvarphi),
\ee
the Green function for $\psi$ can be obtained from $G^y$ by 
inserting the expressions for $y$ and $F$ (Eq. \ref{ke}) into Eq. \ref{reso}
and dividing by $w(\bvarphi)$ 
\begin{equation}
\psi(\bvarphi)=\frac{1}{w(\bvarphi)}\int G^{\psi}_{\lambda}(\bvarphi, \bvarphi_0) 
f(\bvarphi_0)d\bvarphi_0=\frac{1}{w(\bvarphi)}\int G^y_{\lambda}(\bvarphi, \bvarphi_0) 
\Lambda w(\bvarphi_0)f(\bvarphi_0)d\bvarphi_0.
\end{equation}
Collecting all the factors, the resolvent for $\psi(\bvarphi)$ becomes
\begin{equation}
G_{\lambda}^{\psi}(\bvarphi, \bvarphi_0)=\Lambda \frac{w(\bvarphi_0)}{w(\bvarphi)}
 G^y_{\lambda}(\bvarphi, \bvarphi_0).
\end{equation}
 To get the resolvent $G_{\lambda}(g_1g_0^{-1})$
that is used to obtain the state $\psi(g_1)$ on the group manifold
\be
\psi(g_1)=\int_{\sf G}G_\lambda(g_1g_0^{-1})f(g_0)d\mu(g_0)
\ee
two additional steps are necessary. First, we must take into account the different integration
measures in the flat $r$-dimensional space (the torus) and on the group manifold.
According to appendix \ref{app6} the relation between the two integration measures is
\be
V_{{\sf G}}=\int_{{\sf G}}d\mu(g)=V_{{\sf G/T}}\int_{{\sf T}}2^{n-r}\left[w(\bvarphi)\right]^2 d\bvarphi.
\ee
Thus, we have to divide the Green function $G^{\psi}$ by $V_{{\sf G/T}}2^{n-r}\left[w(\bvarphi_0)\right]^2$.
Second, the evolution depends on the radial coordinates $\bvarphi$ of the group
element $g=g_1g_0^{-1}$ that appears in Eq. \ref{kf}, i.e. on the `distance' from the origin
to the point $h\in {\sf T}$. Therefore $\bvarphi_0$ does not have any meaning and
 should be put to zero, while $\bvarphi=\bvarphi(g_1g_0^{-1})$.
The final expression for the resolvent is
\begin{equation}
G_{\lambda}(\bvarphi(g_1g_0^{-1}))=\left.
\frac{\Lambda}{V_{{\sf G/T}}2^{n-r}[w(\bvarphi_0)]^2} G^{\psi}
(\bvarphi,\bvarphi_0)\right|_{\bvarphi_0=0}= \frac{\Lambda}{V_{{\sf G/T}}w(\bvarphi)}\left[
 \frac{G^y(\bvarphi,\bvarphi_0)}{2^{n-r}w(\bvarphi_0)} \right|_{\bvarphi_0=0}.
\end{equation}
(One should note that exactly the same procedure is used when looking for the
Green function in flat space which depends only on the distance between two points
\[
G(\bq,\bq_0)=G(r),\;\;\;r=\sqrt{\bq^2-\bq_0^2}.\] 
$G(r)$ is found by finding the resolvent for the one dimensional radial Laplacean, $G^1(r,r_0)$.
$w(\bvarphi)^2$ plays the role of the part of the integration measure that depends on $r$.
To get $G(r)$ it is necessary to divide $G^1$ 
by the angular volume and by $r_0^{n-1}$ for an $n$
dimensional space, and then set $r_0=0$. The only difference between this case and 
 the group manifold is that there are several radial parameters instead of one).

\section{Bases for the generators of several groups}
\subsection{Rotation generators in spinor representation}
\label{app9}
The  generators of the rotation group SO(p,q)  satisfy the following
commutation relations\cite{dobrev}
\begin{equation}
[L_{ab},L_{cd}]=\delta_{ac}L_{bd}-\delta_{ad}L_{bc}-\delta_{bc}L_{ad}+
\delta_{bd}L_{ac},
\end{equation}
where $\delta_{ab}$ is the metric tensor in the $p+q$ flat space.
The rotation matrices in the 4-dim. spinor representation of
the groups SO(4), SO(5) and SO(6) are built from the Euclidean $\gamma$
matrices:
\begin{eqnarray}
& &L_{ij}=\frac{i}{2}\left[\gamma_j\gamma_i-\gamma_i\gamma_j\right],\;\;\;i,j=1,...,4\nonumber \\
& &L_{i5}=\gamma_i\;\;\;\;i=1,...,4 \nonumber \\
& &L_{i6}=i\gamma_i\gamma_5\;\;\;\;L_{56}=\gamma_5, \;\;\;\;i=1,...,4
\end{eqnarray}
For example, the following basis for the  $\gamma$ matrices can be chosen:
\[ \gamma_j=\left(\begin{array}{cc}0&-i\sigma_j\\i\sigma_j&0\end{array}\right)\;\;\;\;
\gamma_4=\left(\begin{array}{cc}0&-I_2\\-I_2&0\end{array}\right),\;\;\;\;
\gamma_5=\left(\begin{array}{cc}-I_2&0\\0& I_2\end{array}\right).\]
$\sigma_i,\;i=1,2,3$ are the Pauli matrices.

The general method of building the rotation matrices in spinor representation
for any rotation group is explained in a paper by Brauer\cite{brauer}.

\subsection{The generators of $C_3$}
\label{app10}

Choosing the explicit basis of $C_3$ is based on Refs. \cite{cahn,gilmore}.
The general algebra element in this basis is
\be x=\xi^a X_a=\left(\begin{array}{cccccc}
\xi_1&\xi_4+i\xi_5&\xi_6+i\xi_7& \xi_{10}+i\xi_{11} &\xi_{12}+i\xi_{13}
&\xi_{14}+i\xi_{15}\\
\xi_4-i\xi_5&\xi_2&\xi_8+i\xi_9&\xi_{12}+i\xi_{13}&\xi_{16}+i\xi_{17}&\xi_{18}+i\xi_{19}\\
\xi_6-i\xi_7&\xi_8-i\xi_9&\xi_3&\xi_{14}+i\xi_{15} &\xi_{18}+i\xi_{19}&\xi_{20}+i\xi_{21}\\
 \xi_{10}-i\xi_{11} &\xi_{12}-i\xi_{13}&\xi_{14}-i\xi_{15}&-\xi_1&-\xi_4+i\xi_5
&-\xi_6+i\xi_7\\
\xi_{12}-i\xi_{13}&\xi_{16}-i\xi_{17}&\xi_{18}-i\xi_{19}&-\xi_4-i\xi_5&-\xi_2
&-\xi_8+i\xi_9\\
\xi_{14}-i\xi_{15} &\xi_{18}-i\xi_{19}&\xi_{20}-i\xi_{21}&-\xi_6-i\xi_7&-\xi_8-i\xi_9
&-\xi_3
\end{array} \right).\ee

\newpage
\begin{figure}
\begin{center}
\leavevmode
\epsfig{file=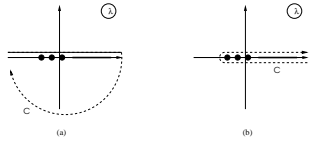,height=4cm,angle=270}
\caption{The integration contour {\sf C} in the complex $\lambda$ plane} 
\end{center}
\end{figure}

\begin{figure}
\begin{center}
\leavevmode
\epsfig{file=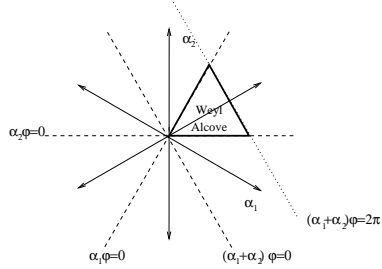,height=5cm,angle=270}
\caption{Root diagram and the Weyl alcove of SU(3)}
\end{center}
\end{figure}

\begin{figure}
\begin{center}
\leavevmode
\epsfig{file=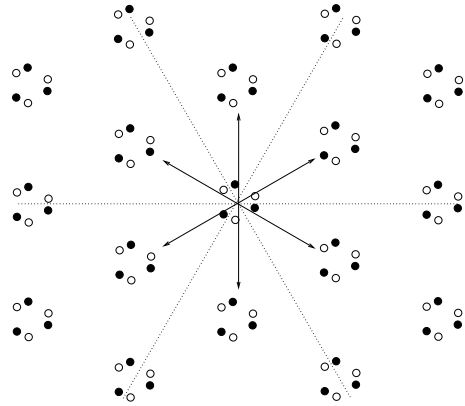,height=6cm,angle=270}
\caption{Images that are created by reflections of 
a point in the Weyl alcove of SU(3).  Full(empty) circles correspond to odd(even)
 reflections }
\end{center}
\end{figure}

\begin{figure}
\begin{center}
\leavevmode
\epsfig{file=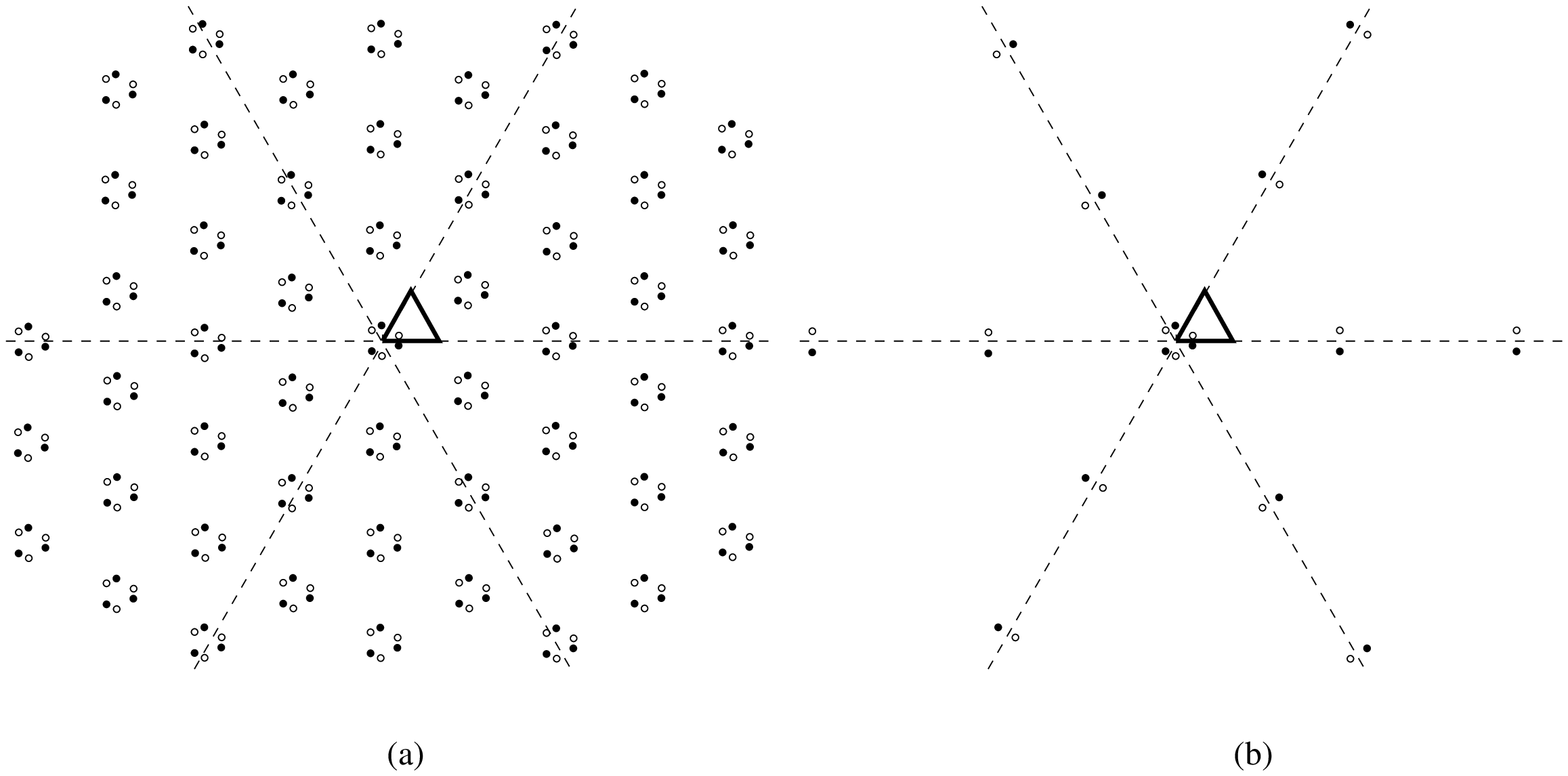,height=7cm,angle=270}
\caption{SU(2,1) has two evolution domains. The equivalent points for 
each domain are shown. Full(empty) circles correspond to odd(even)
 reflections } 
\end{center}
\end{figure}

\begin{figure}
\begin{center}
\leavevmode
\epsfig{file=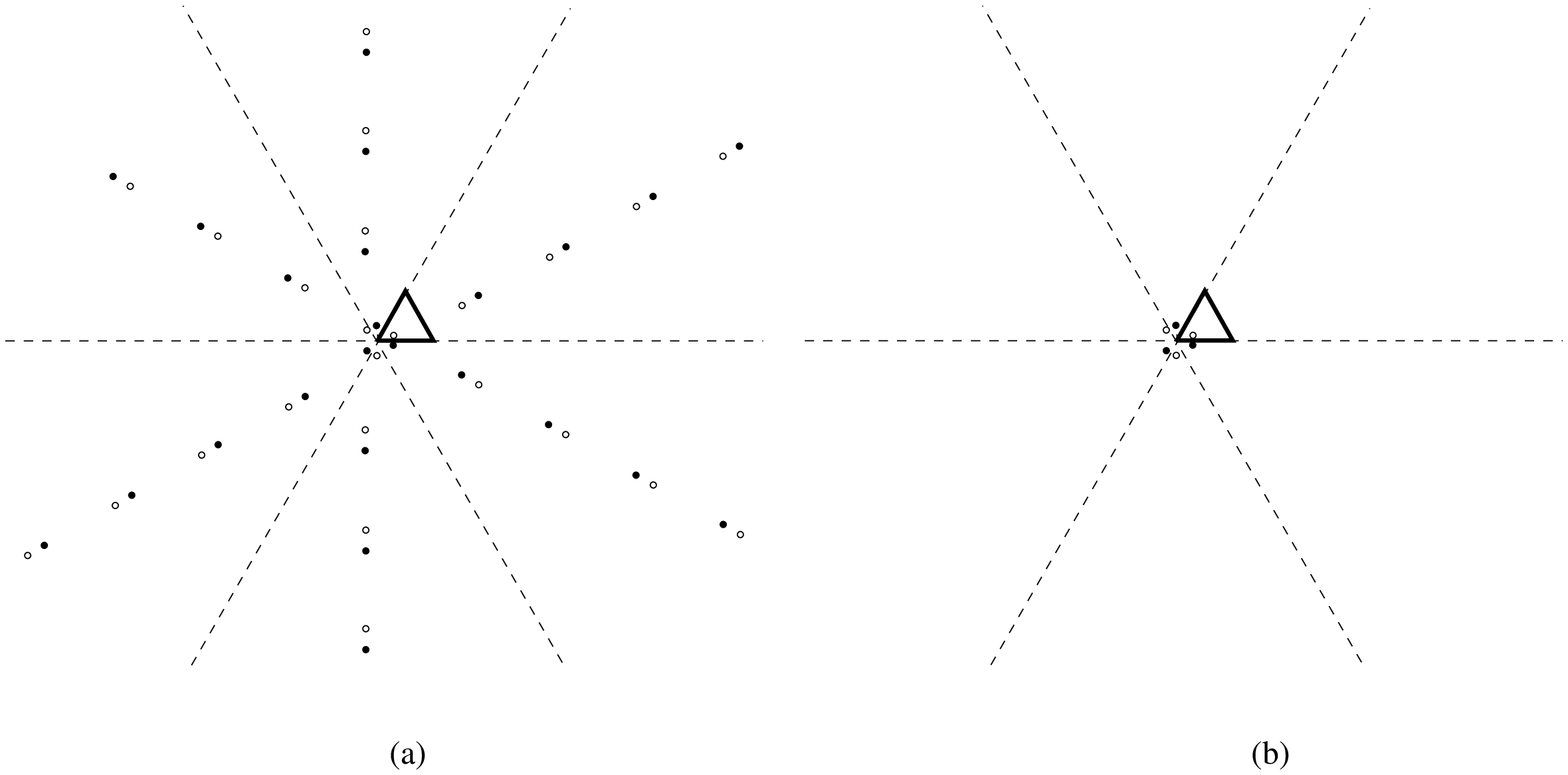,height=7cm,angle=270}
\caption{SL(3,R) has two evolution domains. The equivalent points for 
each domain are shown. Full(empty) circles correspond to odd(even)
 reflections } 
\end{center}
\end{figure}

\begin{thebibliography}{99}
\bibitem{barut1}A.O.Barut and R. Raczka, {\em Proc. Roy. Soc. London} {\bf A287}
(1965) 519%-531.
\bibitem{bellman} R. Bellman, ``A brief introduction to Theta functions"
(Holt, Rinehart \& Wilson, New York, 1961.)
\bibitem{ber56}F. A. Berezin, {\em Math. USSR- Doklady} {\bf 107} (1956) 9.
\bibitem{ber57}F. A. Berezin, {\em Trans. Moscow Math. Soc.} {\bf 6} (1957)
371 ({\em Amer. Math. Soc. Transl.} (2) {\bf 21} (1962) 239); 
{\bf 12} (1963) 453 (Engl. transl. pp. 510-524).
\bibitem{bopp}F. Bopp and R. Haag, {\em Zeit. f. Naturforschung} {\bf 5a}
(1950) 644.
\bibitem{bourbaki}N. Bourbaki, ``Groupes et
algebres de Lie", Chs. 5-7  (Hermann, Paris, 1960).
\bibitem{brauer} R. Brauer and H. Weyl, {\em Amer. J. Math.} {\bf 57} No. 2 (1935) 425.
\bibitem{cahn} R. N. Cahn, ``Semi-simple Lie algebras and their representations'',
(The Benjamin/Cummings Publishing Company, 1984)
\bibitem{camporesi}R. Camporesi, {\em Phys. Rept.} {\bf 196} (1990) 1.
\bibitem{chevalley} C. Chevalley, `` Theory of Lie groups'', (Princeton University press,
1946).
\bibitem{dobrev} V. K. Dobrev {\em et Al.}, ``Harmonic analysis'', Lecture Notes in
Physics Vol. 63, Springer-Verlag, (1977)
\bibitem{dowker70}J. S. Dowker, {\em J. Phys.} {\bf A3} (1970) 451.
\bibitem{dowker71}J. S. Dowker, {\em Ann. Phys. (NY)} {\bf 62} (1971) 361.
\bibitem{eskin}L. D. Eskin, ``Heat transport equation on Lie groups",
in {\em Collection of papers to the memory of N. G. Chebotarev} (Kazan State
University, 1963), pp. 113-132.
\bibitem{fock}
V. A. Fock, {\em Phys. Z. d. Sowjetunion} {\bf 12} (1937) 404.
\bibitem{gant39a} F. Gantmacher, Mat. Sb. {\bf 5} (1939) 101.
\bibitem{gant39b} F. Gantmacher, Mat. Sb. {\bf 5} (1939) 218.
\bibitem{gilmore} R. Gilmore, ``Lie groups, Lie algebras, and some of their
applications'', (John Wiley \& Sons, New York, 1974). 
\bibitem{greiner} W. Greiner and B.Muller, ``Quantum mechanics --- symmetries'',
(Springer-Verlag Berlin, Heidelberg, 1989).
\bibitem{hadamard}J. Hadamard, ``Lectures on Cauchy's problem in linear
partial differential equations" (Dover, New York, 1952).
\bibitem{hermann}R. Hermann, ``Lie groups for physicists'',
(W.A. Bemjamin, Inc. New York, 1966).
\bibitem{marter78}M. S. Marinov and M. V. Terentyev, {\em Sov. J. Nucl. 
Phys.} {\bf 28} (1978) 729.
\bibitem{mlong} M. S. Marinov and M. V. Terentiev,% Dynamics on the Group Manifold
{\em  Fortschritte der Physik} {\bf 27} (1979) 511%--545
\bibitem{minv} M. S. Marinov, %Invariant Volumes of Compact Groups,
{\em  J. Phys.} {\bf A13} (1980) 3357%--3366
\bibitem{morse} P. M. Morse and H. Feshbach, ``Methods of theoretical physics'',
(McGraw-Hill Book Company, 1978)
\bibitem{nambu}Y. Nambu, {\em Prog. Theor. Phys.} {\bf 5} (1950) 82.
\bibitem{schulman}L. Schulman, {\em Phys. Rev.} {\bf 176} (1968) 1558.
\bibitem{schwinger}J. Schwinger, {\em Phys. Rev.} {\bf 82} (1951) 664.
\bibitem{stuckelberg}E. C. G. Stueckelberg, {\em Helv. Phys. Acta}
{\bf 14} (1941) 322, 588.
\bibitem{tits} J. Tits, ``Tabellen zu den einfachen Lie Gruppen und ihren Darstellungen'',
(Springer-Verlag, Berlin, 1967).
\end{thebibliography}
\end{document}